*Research Note*

# Extremal Behaviour in Multiagent Contract Negotiation

**Paul E. Dunne**                                                    PED@CSC.LIV.AC.UK
*Department of Computer Science*
*The University of Liverpool, Liverpool, UK*

## Abstract

We examine properties of a model of resource allocation in which several agents exchange resources in order to optimise their individual holdings. The schemes discussed relate to well-known negotiation protocols proposed in earlier work and we consider a number of alternative notions of "rationality" covering both quantitative measures, e.g. cooperative and individual rationality and more qualitative forms, e.g. Pigou-Dalton transfers. While it is known that imposing particular rationality and structural restrictions may result in some reallocations of the resource set becoming unrealisable, in this paper we address the issue of the number of restricted rational deals that may be required to implement a particular reallocation when it *is* possible to do so. We construct examples showing that this number may be exponential (in the number of resources $m$), even when all of the agent utility functions are monotonic. We further show that $k$ agents may achieve in a single deal a reallocation requiring exponentially many rational deals if at most $k-1$ agents can participate, this same reallocation being unrealisable by any sequences of rational deals in which at most $k-2$ agents are involved.

## 1. Introduction

Mechanisms for negotiating allocation of resources within a group of agents form an important body of work within the study of multiagent systems. Typical abstract models derive from game-theoretic perspectives in economics and among the issues that have been addressed are strategies that agents use to obtain a particular subset of the resources available, e.g. (Kraus, 2001; Rosenschein & Zlotkin, 1994; Sandholm, 1999), and protocols by which the process of settling upon some allocation of resources among the agents involved is agreed, e.g. (Dignum & Greaves, 2000; Dunne, 2003; Dunne & McBurney, 2003; McBurney et al., 2002).

The setting we are concerned with is encapsulated in the following definition.

**Definition 1** *A resource allocation setting is defined by a triple $\langle \mathcal{A}, \mathcal{R}, \mathcal{U} \rangle$ where*

$$\mathcal{A} = \{A_1, A_2, \ldots, A_n\} \quad ; \quad \mathcal{R} = \{r_1, r_2, \ldots, r_m\}$$

*are, respectively, a set of (at least two) agents and a collection of (non-shareable) resources. A utility function, $u$, is a mapping from subsets of $\mathcal{R}$ to rational values. Each agent $A_i \in \mathcal{A}$ has associated with it a particular utility function $u_i$, so that $\mathcal{U}$ is $\langle u_1, u_2, \ldots, u_n \rangle$. An allocation $P$ of $\mathcal{R}$ to $\mathcal{A}$ is a partition $\langle P_1, P_2, \ldots, P_n \rangle$ of $\mathcal{R}$. The value $u_i(P_i)$ is called the* utility *of the resources assigned to $A_i$. A utility function, $u$, is* monotone *if whenever $S \subseteq T$ it holds that $u(S) \leq u(T)$, i.e. the value assigned by $u$ to any set of resources, $T$, is never less than the value $u$ attaches to any subset, $S$ of $T$.*

Two major applications in which the abstract view of Definition 1 has been exploited are e-commerce and distributed task realisation. In the first $\mathcal{R}$ represents some collection of commodities offered for sale and individual agents seek to acquire a subset of these, the "value" an agent attaches to a specific set being described by that agent's utility function. In task planning, the "resource" set describes a collection of sub-tasks to be performed in order to realise some complex task, e.g. the "complex task" may be to transport goods from a central warehouse to some set of cities. In this example $\mathcal{R}$ describes the locations to which goods must be dispatched and a given allocation defines those places to which an agent must arrange deliveries. The utility functions in such cases model the cost an agent associates with carrying out its alloted sub-tasks.

Within the very general context of Definition 1, a number of issues arise stemming from the observation that it is unlikely that some *initial allocation* will be seen as satisfactory either with respect to the views of all agents in the system or with respect to divers global considerations. Thus, by proposing changes to the initial assignment individual agents seek to obtain a "better" allocation. This scenario raises two immediate questions: how to evaluate a given partition and thus have a basis for forming improved or optimal allocations; and, the issue underlying the main results of this paper, what restrictions should be imposed on the form that proposed deals may take.

We shall subsequently review some of the more widely studied approaches to defining conditions under which some allocations are seen as "better" than others. For the purposes of this introduction we simply observe that such criteria may be either *quantitative* or *qualitative* in nature. As an example of the former we have the approach wherein the "value" of an allocation $P$ is simply the sum of the values given by the agents' utility functions to the subsets of $\mathcal{R}$ they have been apportioned within $P$, i.e. $\sum_{i=1}^{n} u_i(P_i)$: this is the so-called *utilitarian social welfare*, which to avoid repetition we will denote by $\sigma_u(P)$. A natural aim for agents within a commodity trading context is to seek an allocation under which $\sigma_u$ is *maximised*. One example of a *qualitative* criterion is "*envy freeness*": informally, an allocation, $P$, is envy-free if no agent assigns greater utility to the resource set $(P_j)$ held by another agent than it does with respect to the resource set $(P_i)$ it has actually been allocated, i.e. for each distinct pair $\langle i, j \rangle$, $u_i(P_i) \geq u_i(P_j)$.

In very general terms there are two approaches that have been considered in treating the question of how a finite collection of resources might be distributed among a set of agents in order to optimise some criterion of interest: "contract-net" based methods, e.g. (Dunne et al., 2003; Endriss et al., 2003; Endriss & Maudet, 2004b; Sandholm, 1998, 1999) deriving from the work of Smith (1980); and "combinatorial auctions", e.g. (Parkes & Ungar, 2000a, 2000b; Sandholm et al., 2001; Sandholm, 2002; Sandholm & Suri, 2003; Tennenholz, 2000; Yokoo et al., 2004, amongst others). The significant difference between these is in the extent to which a centralized controlling agent determines the eventual distribution of resources among agents.

One may view the strategy underlying combinatorial auctions as investing the computational effort into a "pre-processing" stage following which a given allocation is determined. Thus a controlling agent (the "auctioneer") is supplied with a set of *bids* – pairs $\langle S_j, p_j \rangle$ wherein $S_j$ is some subset of the available resources and $p_j$ the price agent $A_j$ is prepared to pay in order to acquire $S_j$. The problem faced by the auctioneer is to decide which bids to accept in order to maximise the overall profit subject to the constraint that each item can be obtained by at most one agent.

What we shall refer to as "contract-net schemes" typically eschew the precomputation stage and subordination to a controlling arbiter employed in auction mechanisms, seeking instead to realise a suitable allocation by an agreed sequence of deals. The *contract-net* (in its most general instantiation) for scenarios of $m$ resources distributed among $n$ agents is the complete directed graph with $n^m$ vertices (each of which is associated with a distinct allocation). In this way a possible deal $\langle P, Q \rangle$ is represented as an edge directed from the vertex labelled with $P$ to that labelled $Q$. Viewed thus, identifying a sequence of deals can be interpreted as a search process which, in principle, individual agents may conduct in an autonomous fashion.

Centralized schemes can be effective in contexts where the participants cooperate (in the sense of accepting the auctioneer's arbitration). In environments within which agents are highly self-interested to the extent that their aims conflict with the auction process or in which there is a high degree of "uncertainty" about the outcome, in working towards a final allocation, the agents involved may only be prepared to proceed "cautiously": that is, an agent will only accept a proposed reallocation if satisfied that such would result in an *immediate* improvement from its own perspective. In such cases, the process of moving from the initial allocation, $P_{init}$, to the eventual reallocation $P_{fin}$ is by a sequence of local *rational* deals, e.g. an agent might refuse to accept deals which reduced $\sigma_u$ because of the possibility that it suffers an uncompensated loss in utility. A key issue here is the following: if the deal protocol allows only moves in which at each stage some agent $A_j$ offers a single resource to another agent $A_j$ then the rational reallocation $\langle P_{init}, P_{fin} \rangle$ can *always* be implemented; if, however, every single move must be "rational" then $\langle P_{init}, P_{fin} \rangle$ may not be realisable.

We may, informally, regard the view of such agents as "myopic", in the sense that they are unwilling to accept a "short-term loss" (a deal $\langle P, Q \rangle$ under they might incur a loss of utility) despite the prospect of a "long-term gain" (assuming $\sigma_u(P_{fin}) > \sigma_u(P_{init})$ holds).

There are a number of reasons why an agent may adopt such views, e.g. consider the following simple protocol for agreeing a reallocation.

> A reallocation of resources is agreed over a sequence of stages, each of which involves communication between two agents, $A_i$ and $A_j$. This communication consists of $A_i$ issuing a proposal to $A_j$ of the form $(buy, r, p)$, offering to purchase $r$ from $A_j$ for a payment of $p$; or $(sell, r, p)$, offering to transfer $r$ to $A_j$ in return for a payment $p$. The response from $A_j$ is simply *accept* (following which the deal is implemented) or *reject*.

This, of course, is a very simple negotiation structure, however consider its operation within a two agent setting in which one agent, $A_1$ say, wishes to bring about an allocation $P_{fin}$ (and thus can devise a plan – sequence of deals – to realise this from an initial allocation $P_{init}$) while the other agent, $A_2$, does *not* know $P_{fin}$. In addition, assume that $A_1$ is the only agent that makes proposals and that a final allocation is fixed either when $A_1$ is "satisfied" or as soon as $A_2$ *rejects* any offer.

While $A_2$ *could* be better off if $P_{fin}$ is realised, it may be the case that the only proposals $A_2$ will accept are those under which it does not lose, e.g. some agents may be sceptical about the *bona fides* of others and will accept only deals from which they can perceive an immediate benefit. There are several reasons why an agent may embrace such attitudes within the schema outlined: once a deal has been implemented $A_2$ may lose utility but no further proposals are made by $A_1$ so that the loss is "permanent". We note that even if we enrich the basic protocol so that $A_1$ can describe $P_{fin}$, $A_2$ may still reject offers under which it suffers a loss, since it is unwilling to rely on the subsequent deals that would ameliorate its loss actually being proposed. Although the position taken by $A_2$ in the setting just described may appear unduly cautious, we would claim that it does reflect "real" behaviour in certain contexts. Outside the arena of automated allocation and negotiation in multiagent systems, there are many examples of actions by individuals where promised long-term gains are insufficient to engender the acceptance of short term loss. Consider "chain letter" schemes (or their more subtle manifestation as "pyramid selling" enterprises): such have a natural lifetime bounded by the size of the population in which they circulate, but may break down before this is reached. Faced with a request to "send $10 to the five names at the head of the list and forward the letter to ten others after adding your name" despite the possibility of significant gain after a temporary loss of $50, to ignore such blandishments is not seen as overly sceptical and cautious: there may be reluctance to accept that one will eventually receive sufficient recompense in return and suspicion that the name order has been manipulated.

In summary, we can identify two important influences that lead to contexts in which agents prefer to move towards a reallocation via a sequence of "rational" deals. Firstly, the agents are self-interested but operating in an unstable environment, e.g. in the "chain letter" setting, an agent cannot reliably predict the exact point at which the chain will fail. The second factor is that computational restrictions may limit the decisions an individual agent can make about whether or not to accept a proposed deal. For example in settings where all deals involve one resource at a time, $A_2$ may reject a proposal to accept some resource, $r$, since $r$ is only "useful" following a further sequence of deals: if this number of further deals is "small" then $A_2$ could decide to accept the proposed deal since it has sufficient computational power to determine that there is a context in which $r$ is of value; if this number is "large" however, then $A_2$ may lack sufficient power to scan the search space of future possibilities that would allow it to accept $r$. Notice that in the extreme case, $A_2$ makes its decision solely on whether $r$ is of immediate use, i.e. $A_2$ is myopic. A more powerful $A_2$ may be able to consider whether $r$ is useful should up to $k$ further deals take place: in this case, $A_2$ could still refuse to accept $r$ since, although of use, $A_2$ cannot determine this with a bounded look ahead.

In total for the scenario we have described, if $A_1$ wishes to bring about an allocation $P_{fin}$ then faced with the view adopted by $A_2$ and the limitations imposed by the deal protocol, the only "effective plan" that $A_1$ could adopt is to find a sequence of *rational* deals to propose to $A_2$.

Our aim in this article is to show that combining "structural" restrictions (e.g. only one resource at a time is involved in a local reallocation) with rationality restrictions can result in settings in which any sequence to realise a reallocation $\langle P, Q \rangle$ must involve exponentially many (in $|\mathcal{R}|$) separate stages. We refine these ideas in the next sub-section.

## 1.1 Preliminary Definitions

To begin, we first formalise the concepts of *deal* and *contract path*.

**Definition 2** *Let $\langle \mathcal{A}, \mathcal{R}, \mathcal{U} \rangle$ be a resource allocation setting. A* deal *is a pair $\langle P, Q \rangle$ where $P = \langle P_1, \ldots, P_n \rangle$ and $Q = \langle Q_1, \ldots, Q_n \rangle$ are distinct partitions of $\mathcal{R}$. The effect of implementing the deal $\langle P, Q \rangle$ is that the allocation of resources specified by $P$ is replaced with that specified by $Q$. Following the notation of (Endriss & Maudet, 2004b) for a deal $\delta = \langle P, Q \rangle$, we use $\mathcal{A}^\delta$ to indicate the subset of $\mathcal{A}$ involved, i.e. $A_k \in \mathcal{A}^\delta$ if and only if $P_k \neq Q_k$.*

*Let $\delta = \langle P, Q \rangle$ be a deal. A* contract path *realising $\delta$ is a sequence of allocations*

$$\Delta = \langle P^{(1)}, \ P^{(2)}, \ldots, \ P^{(t-1)}, \ P^{(t)} \rangle$$

*in which $P = P^{(1)}$ and $P^{(t)} = Q$. The* length *of $\Delta$, denoted $|\Delta|$ is $t - 1$, i.e. the number of deals in $\Delta$.*

There are two methods which we can use to reduce the number of deals that a single agent may have to consider in seeking to move from some allocation to another, thereby avoiding the need to choose from exponentially many alternatives: *structural* and *rationality* constraints. Structural constraints limit the permitted deals to those which bound the number of resources and/or the number of agents involved, but take no consideration of the view any agent may have as to whether its allocation has improved. In contrast, rationality constraints restrict deals $\langle P, Q \rangle$ to those in which $Q$ "improves" upon $P$ according to particular criteria. In this article we consider two classes of structural constraint: *O*-contracts, defined and considered in (Sandholm, 1998), and what we shall refer to as $M(k)$-contracts.

**Definition 3** *Let $\delta = \langle P, Q \rangle$ be a deal involving a reallocation of $\mathcal{R}$ among $\mathcal{A}$.*

*a. $\delta$ is a* one contract *(O-contract) if*

    *O1. $\mathcal{A}^\delta = \{i, j\}$.*

    *O2. There is a unique resource $r \in P_i \cup P_j$ for which $Q_i = P_i \cup \{r\}$ and $Q_j = P_j \setminus \{r\}$ (with $r \in P_j$) or $Q_j = P_j \cup \{r\}$ and $Q_i = P_i \setminus \{r\}$ (with $r \in P_i$)*

*b. For a value $k \geq 2$, the deal $\delta = \langle P, Q \rangle$ is an $M(k)$-contract if $2 \leq |\mathcal{A}^\delta| \leq k$ and $\cup_{i \in \mathcal{A}^\delta} Q_i = \cup_{i \in \mathcal{A}^\delta} P_i$.*

Thus, *O*-contracts involve the transfer of *exactly one* resource from a particular agent to another, resulting in the number of deals compatible with any given allocation being exactly $(n-1)m$: each of the $m$ resources can be reassigned from its current owner to any of the other $n - 1$ agents.

Rationality constraints arise in a number of different ways. For example, from the standpoint of an individual agent $A_i$ a given deal $\langle P, Q \rangle$ may have three different outcomes: $u_i(P_i) < u_i(Q_i)$, i.e. $A_i$ values the allocation $Q_i$ as superior to $P_i$; $u_i(P_i) = u_i(Q_i)$, i.e. $A_i$ is indifferent between $P_i$ and $Q_i$; and $u_i(P_i) > u_i(Q_i)$, i.e. $A_i$ is worse off after the deal. When global optima such as utilitarian social welfare are to be maximised, there is the question of what incentive there is for any agent to accept a deal $\langle P, Q \rangle$ under which it is left with a less valuable resource holding. The standard approach to this latter question is to introduce the notion of a *pay-off* function, i.e. in order for $A_i$ to accept a deal under which it suffers a reduction in utility, $A_i$ receives some payment sufficient to compensate for its loss. Of course such compensation must be made by other agents in the system who in providing it do not wish to pay in excess of any gain. In defining notions of pay-off the interpretation is that in any transaction each agent $A_i$ makes a payment, $\pi_i$: if $\pi_i < 0$ then $A_i$ is given $-\pi_i$ in return for accepting a deal; if $\pi_i > 0$ then $A_i$ contributes $\pi_i$ to the amount to be distributed among those agents whose pay-off is negative.

This notion of "sensible transfer" is captured by the concept of *individual rationality*, and is often defined in terms of an appropriate pay-off vector existing. It is not difficult, however, to show that such definitions are equivalent to the following.

**Definition 4** *A deal $\langle P, Q \rangle$ is* individually rational *(IR) if and only if $\sigma_u(Q) > \sigma_u(P)$.*

We shall consider alternative bases for rationality constraints later: these are primarily of interest within so-called *money free* settings (so that compensatory payment for a loss in utility is not an option).

The central issue of interest in this paper concerns the properties of the contract-net graph when the allowed deals must satisfy both a structural *and* a rationality constraint. Thus, if we consider arbitrary predicates $\Phi$ on deals $\langle P, Q \rangle$ – where the cases of interest are $\Phi$ combining a structural and rationality condition – we have,

**Definition 5** *For $\Phi$ a predicate over distinct pairs of allocations, a contract path*

$$\langle P^{(1)}, \ P^{(2)}, \ldots, \ P^{(t-1)}, \ P^{(t)} \rangle$$

*realising $\langle P, Q \rangle$ is a $\Phi$-path if for each $1 \leq i < t$, $\langle P^{(i)}, P^{(i+1)} \rangle$ is a $\Phi$-deal, that is $\Phi(P^{(i)}, P^{(i+1)})$ holds. We say that $\Phi$ is* complete *if any deal $\delta$ may be realised by a $\Phi$-path. We, further, say that $\Phi$ is* complete with respect to $\Psi$-deals *(where $\Psi$ is a predicate over distinct pairs of allocations) if any deal $\delta$ for which $\Psi(\delta)$ holds may be realised by a $\Phi$-path.*

The main interest in earlier studies of these ideas has been in areas such as identifying necessary and/or sufficient conditions on deals to be complete with respect to particular criteria, e.g. (Sandholm, 1998); and in establishing "convergence" and termination properties, e.g. Endriss et al. (2003), Endriss and Maudet (2004b) consider deal types, $\Phi$, such that every maximal[1] $\Phi$-path ends in a Pareto optimal allocation, i.e. one in which any reallocation under which some agent improves its utility will lead to another agent suffering a loss. Sandholm (1998) examines how restrictions e.g. with $\Phi(P, Q) = \top$ if and only if $\langle P, Q \rangle$ is an *O*-contract, may affect the existence of contract paths to realise deals. Of particular interest, from the viewpoint of heuristics for exploring the contract-net graph, are cases where $\Phi(P, Q) = \top$ if and only if the deal $\langle P, Q \rangle$ is individually rational. For the case of *O*-contracts the following are known:

**Theorem 1**

    *a. O-contracts are complete.*

---

1. "Maximal" in the sense that if $\langle P^{(1)}, \ldots, P^{(t)} \rangle$ is such a path, then for every allocation, $Q$, $\Phi(P^{(t)}, Q)$ does not hold.

     b. *IR O-contracts are* not *complete with respect to IR deals.*

In the consideration of algorithmic and complexity issues presented in (Dunne et al., 2003) one difficulty with attempting to formulate reallocation plans by rational $O$-contracts is already apparent, that is:

**Theorem 2** *Even in the case $n = 2$ and with monotone utility functions the problem of deciding if an IR O-contract path exists to realise the IR deal $\langle P, Q \rangle$ is* NP–*hard.*

Thus deciding if any rational plan is possible is already computationally hard. In this article we demonstrate that, even if an appropriate rational plan exists, in extreme cases, there may be significant problems: the number of deals required could be exponential in the number of resources, so affecting both the time it will take for the schema outlined to conclude and the space that an agent will have to dedicate to storing it. Thus in his proof of Theorem 1 (b), Sandholm observes that when an IR $O$-contract path exists for a given IR deal, it may be the case that its length exceeds $m$, i.e. some agent passes a resource to another and then accepts the same resource at a later stage.

    The typical form of the results that we derive can be summarised as:

> For $\Phi$ a structural constraint ($O$-contract or $M(k)$-contract) and $\Psi$ a rationality constraint, e.g. $\Psi(P, Q)$ holds if $\langle P, Q \rangle$ is individually rational, there are resource allocation settings $\langle \mathcal{A}_n, \mathcal{R}_m, \mathcal{U} \rangle$ in which there is a deal $\langle P, Q \rangle$ satisfying all of the following.
>
>     a. $\langle P, Q \rangle$ is a $\Psi$-deal.
>
>     b. $\langle P, Q \rangle$ can be realised by a contract path on which every deal satisfies the structural constraint $\Phi$ *and* the rationality constraint $\Psi$.
>
>     c. Every such contract path has length at least $g(m)$.

For example, we show that there are instances for which the shortest IR $O$-contract path has length exponential in $m$.[2] In the next section we will be interested in lower bounds on the values of the following functions: we introduce these in general terms to avoid unnecessary subsequent repetition.

**Definition 6** *Let $\langle \mathcal{A}, \mathcal{R}, \mathcal{U} \rangle$ be a resource allocation setting. Additionally let $\Phi$ and $\Psi$ be two predicates on deals. For a deal $\delta = \langle P, Q \rangle$ the partial function $L^{\mathrm{opt}}(\delta, \langle \mathcal{A}, \mathcal{R}, \mathcal{U} \rangle, \Phi)$ is the length of the shortest $\Phi$-contract path realising $\langle P, Q \rangle$ if such a path exists (and is undefined if no such path is possible). The partial function $L^{\mathrm{max}}(\langle \mathcal{A}, \mathcal{R}, \mathcal{U} \rangle, \Phi, \Psi)$ is*

$$L^{\mathrm{max}}(\langle \mathcal{A}, \mathcal{R}, \mathcal{U} \rangle, \Phi, \Psi) \;=\; \max_{\Psi\text{-}deals\ \delta}\ L^{\mathrm{opt}}(\delta, \langle \mathcal{A}, \mathcal{R}, \mathcal{U} \rangle, \Phi)$$

*Finally, the partial function $\rho^{\mathrm{max}}(n, m, \Phi, \Psi)$ is*

$$\rho^{\mathrm{max}}(n, m, \Phi, \Psi) \;=\; \max_{\mathcal{U} = \langle u_1, u_2, \ldots, u_n \rangle}\ L^{\mathrm{max}}(\langle \mathcal{A}_n, \mathcal{R}_m, \mathcal{U} \rangle, \Phi, \Psi)$$

*where consideration is restricted to those $\Psi$-deals $\delta = \langle P, Q \rangle$ for which a realising $\Phi$-path exists.*

---

2. Sandholm (1998) gives an upper bound on the length of such paths which is also exponential in $m$, but does not explicitly state any lower bound other than that already referred to.

The three measures, $L^{\mathrm{opt}}$, $L^{\mathrm{max}}$ and $\rho^{\mathrm{max}}$ distinguish different aspects regarding the length of contract-paths. The function $L^{\mathrm{opt}}$ is concerned with $\Phi$-paths realising a single deal $\langle P, Q \rangle$ in a given resource allocation setting $\langle \mathcal{A}, \mathcal{R}, \mathcal{U} \rangle$: the property of interest being the number of deals in the *shortest*, i.e. optimal length, $\Phi$-path. We stress that $L^{\mathrm{opt}}$ is a *partial* function whose value is undefined in the event that $\langle P, Q \rangle$ cannot be realised by a $\Phi$-path in the setting $\langle \mathcal{A}, \mathcal{R}, \mathcal{U} \rangle$. The function $L^{\mathrm{max}}$ is defined in terms of $L^{\mathrm{opt}}$, again in the context of a specific resource allocation setting. The behaviour of interest for $L^{\mathrm{max}}$, however, is not simply the length of $\Phi$-paths realising a specific $\langle P, Q \rangle$ but the "worst-case" value of $L^{\mathrm{opt}}$ for deals which are $\Psi$-deals. We note the qualification that $L^{\mathrm{max}}$ is defined only for $\Psi$-deals that *are* capable of being realised by $\Phi$-paths, and thus do not consider cases for which no appropriate contract path exists. Thus, if it should be the case that no $\Psi$-deal in the setting $\langle \mathcal{A}, \mathcal{R}, \mathcal{U} \rangle$ can be realised by a $\Phi$-path then the value $L^{\mathrm{max}}(\langle \mathcal{A}, \mathcal{R}, \mathcal{U} \rangle, \Phi, \Psi)$ is undefined, i.e. $L^{\mathrm{max}}$ is also a partial function. We may interpret any *upper* bound on $L^{\mathrm{max}}$ in the following terms: if $L^{\mathrm{max}}(\langle \mathcal{A}, \mathcal{R}, \mathcal{U} \rangle, \Phi, \Psi) \leq K$ then any $\Psi$-deal *for which a $\Phi$-path exists* can be realised by a $\Phi$-path of length at most $K$.

Our main interest will centre on $\rho^{\mathrm{max}}$ which is concerned with the behaviour of $L^{\mathrm{max}}$ as a function of $n$ and $m$ and ranges over *all $n$-tuples* of utility functions $\langle u : 2^{\mathcal{R}} \to \mathbf{Q} \rangle^n$. Our approach to obtaining lower bounds for this function is *constructive*, i.e. for each $\langle \Phi, \Psi \rangle$ that is considered, we show how the utility functions $\mathcal{U}$ may be defined in a setting with $m$ resources so as to yield a lower bound on $\rho^{\mathrm{max}}(n, m, \Phi, \Psi)$. In contrast to the measures $L^{\mathrm{opt}}$ and $L^{\mathrm{max}}$, the function $\rho^{\mathrm{max}}$ is not described in terms of a single fixed resource allocation setting. It is, however, still a *partial* function: depending on $\langle n, m, \Phi, \Psi \rangle$ it may be the case that in *every $n$* agent, $m$ resource allocation setting, regardless of which choice of utility functions is made, there is no $\Psi$-deal, $\langle P, Q \rangle$ capable of being realised by $\Phi$-path, and for such cases the value of $\rho^{\mathrm{max}}(n, m, \Phi, \Psi)$ will be undefined.[3]

It is noted, at this point, that the definition of $\rho^{\mathrm{max}}$ allows *arbitrary* utility functions to be employed in constructing "worst-case" instances. While this is reasonable in terms of general lower bound results, as will be apparent from the given constructions the utility functions actually employed are highly artificial (and unlikely to feature in "real" application settings). We shall attempt to address this objection by further considering bounds on the following variant of $\rho^{\mathrm{max}}$:

$$\rho^{\mathrm{max}}_{\mathrm{mono}}(n, m, \Phi, \Psi) = \max_{\mathcal{U} = \langle u_1, u_2, \ldots, u_n \rangle \,:\, \text{each } u_i \text{ is monotone}} L^{\mathrm{max}}(\langle \mathcal{A}_n, \mathcal{R}_m, \mathcal{U} \rangle, \Phi, \Psi)$$

Thus, $\rho^{\mathrm{max}}_{\mathrm{mono}}$ deals with resource allocation settings within which all of the utility functions must satisfy a monotonicity constraint.

The main results of this article are presented in the next sections. We consider two general classes of contract path: $O$-contract paths under various rationality conditions in

---

3. In recognising the *possibility* that $\rho^{\mathrm{max}}(n, m, \Phi, \Psi)$ could be undefined, we are *not* claiming that such behaviour arises with any of the instantiations of $\langle \Phi, \Psi \rangle$ considered subsequently: in fact it will be clear from the constructions that, denoting by $\rho^{\mathrm{max}}_{\Phi, \Psi}(n, m)$ the function $\rho^{\mathrm{max}}(n, m, \Phi, \Psi)$ for a fixed instantiation of $\langle \Phi, \Psi \rangle$, with the restricted deal types and rationality conditions examined, the function $\rho^{\mathrm{max}}_{\Phi, \Psi}(n, m)$ is a *total* function. Whether it is possible to formulate "sensible" choices of $\langle \Phi, \Psi \rangle$ with which $\rho^{\mathrm{max}}_{\Phi, \Psi}(n, m)$ is undefined for some values of $\langle n, m \rangle$ (and, if so, demonstrating examples of such) is, primarily, only a question of combinatorial interest, whose development is not central to the concerns of the current article.

Section 2; and, similarly, $M(k)$-contract paths for arbitrary values of $k \geq 2$ in Section 3. Our results are concerned with the construction of resource allocation settings $\langle \mathcal{A}, \mathcal{R}_m, \mathcal{U} \rangle$ for which given some rationality requirement, e.g. that deals be individually rational, there is some deal $\langle P, Q \rangle$ that satisfies the rationality condition, can be realised by a rational $O$-contract path (respectively, $M(k)$-contract path), but with the number of deals required by such paths being exponential in $m$. We additionally obtain slightly weaker (but still exponential) lower bounds for rational $O$-contract paths within settings of monotone utility functions, i.e. for the measure $\rho_{\mathrm{mono}}^{\max}$, outlining how similar results may be derived for $M(k)$-contract paths.

In the resource allocation settings constructed for demonstrating these properties with $M(k)$-contract paths, the constructed deal $\langle P, Q \rangle$ is realisable with a *single* $M(k+1)$-contract but unrealisable by any *rational* $M(k-1)$-contract path. We discuss related work, in particular the recent study of (Endriss & Maudet, 2004a) that addresses similar issues to those considered in the present article, in Section 4. Conclusions and some directions for further work are presented in the final section.

## 2. Lower Bounds on Path Length – $O$-contracts

In this section we consider the issue of contract path length when the structural restriction requires individual deals to be $O$-contracts. We first give an overview of the construction method, with the following subsections analysing the cases of unrestricted utility functions and, subsequently, *monotone* utility functions.

### 2.1 Overview

The strategy employed in proving our results involves two parts: for a given class of restricted contract paths we proceed as follows in obtaining lower bounds on $\rho^{\max}(n, m, \Phi, \Psi)$.

a. For the contract-net graph partitioning $m$ resources among $n$ agents, construct a path, $\Delta_m = \langle P^{(1)}, P^{(2)}, \ldots, P^{(t)} \rangle$ realising a deal $\langle P^{(1)}, P^{(t)} \rangle$. For the *structural* constraint, $\Phi'$ influencing $\Phi$ it is then proved that:

   a1. The contract path $\Delta_m$ *is* a $\Phi'$-path, i.e. for each $1 \leq i < t$, the deal $\langle P^{(i)}, P^{(i+1)} \rangle$ satisfies the structural constraint $\Phi'$.

   a2. For any pair of allocations $P^{(i)}$ and $P^{(i+j)}$ occurring in $\Delta_m$, if $j \geq 2$ then the deal $\langle P^{(i)}, P^{(i+j)} \rangle$ is *not* a $\Phi'$-deal.

   Thus (a1) ensures that $\Delta_m$ is a suitable contract path, while (a2) will guarantee that there is exactly one allocation, $P^{(i+1)}$, that can be reached *within* $\Delta_m$ from any given allocation $P^{(i)}$ in $\Delta_m$ by means of a $\Phi'$-deal.

b. Define utility functions $\mathcal{U}_n = \langle u_1, \ldots, u_n \rangle$ with the following properties

   b1. The deal $\langle P^{(1)}, P^{(t)} \rangle$ is a $\Psi$-deal.

   b2. For the *rationality* constraint, $\Phi''$ influencing $\Phi$, every deal $\langle P^{(i)}, P^{(i+1)} \rangle$ is a $\Phi''$-deal.

b3. For every allocation $P^{(i)}$ in the contract path $\Delta$ and every allocation $Q$ other than $P^{(i+1)}$ the deal $\langle P^{(i)}, Q \rangle$ is *not* a $\Phi$-deal, i.e. it violates either the stuctural constraint $\Phi'$ or the rationality constraint $\Phi''$.

Thus, (a1) and (b2) ensure that $\langle P^{(1)}, P^{(t)} \rangle$ has a defined value with respect to the function $L^{\text{opt}}$ for the $\Psi$-deal $\langle P^{(1)}, P^{(t)} \rangle$, i.e. a $\Phi$-path realising the deal is possible. The properties given by (a2) and (b3) indicate that (within the constructed resource allocation setting) the path $\Delta_m$ is the *unique* $\Phi$-path realising $\langle P^{(1)}, P^{(t)} \rangle$. It follows that $t-1$, the length of this path, gives a *lower bound* on the value of $L^{\text{max}}$ and hence a lower bound on $\rho^{\text{max}}(n, m, \Phi, \Psi)$.

Before continuing it will be useful to fix some notational details.

We use $\mathcal{H}_m$ to denote the $m$-dimensional hypercube. Interpreted as a directed graph, $\mathcal{H}_m$ has $2^m$ vertices each of which is identified with a distinct $m$-bit label. Using $\alpha = a_1 a_2 \dots a_m$ to denote an arbitrary such label, the edges of $\mathcal{H}_m$ are formed by

$$\{ \langle \alpha, \beta \rangle : \alpha \text{ and } \beta \text{ differ in } \textit{exactly one bit position} \}$$

We identify $m$-bit labels $\alpha = a_1 a_2 \dots a_m$ with subsets $S^\alpha$ of $\mathcal{R}_m$, via $r_i \in S^\alpha$ if and only if $a_i = 1$. Similarly, any subset $S$ of $\mathcal{R}$ can be described by a binary word, $\beta(S)$, of length $m$, i.e. $\beta(S) = b_1 b_2 \dots b_m$ with $b_i = 1$ if and only if $r_i \in S$. For a label $\alpha$ we use $|\alpha|$ to denote the number of bits with value 1, so that $|\alpha|$ is the size of the subset $S^\alpha$. If $\alpha$ and $\beta$ are $m$-bit labels, then $\alpha\beta$ is a $2m$-bit label, so that if $\mathcal{R}_m$ and $\mathcal{T}_m$ are disjoint sets, then $\alpha\beta$ describes the union of the subset $S^\alpha$ of $\mathcal{R}_m$ with the subset $S^\beta$ of $\mathcal{T}_m$. Finally if $\alpha = a_1 a_2 \dots a_m$ is an $m$-bit label then $\overline{\alpha}$ denotes the label formed by changing all 0 values in $\alpha$ to 1 and *vice versa*. In this way, if $S^\alpha$ is the subset of $\mathcal{R}_m$ described by $\alpha$ then $\overline{\alpha}$ describes the set $\mathcal{R}_m \setminus S^\alpha$. To avoid an excess of superscripts we will, where no ambiguity arises, use $\alpha$ both to denote the $m$-bit label and the subset of $\mathcal{R}_m$ described by it, e.g. we write $\alpha \subset \beta$ rather than $S^\alpha \subset S^\beta$.

For $n = 2$ the contract-net graph induced by $O$-contracts can be viewed as the $m$-dimensional hypercube $\mathcal{H}_m$: the $m$-bit label, $\alpha$ associated with a vertex of $\mathcal{H}_m$ describing the allocation $\langle \alpha, \overline{\alpha} \rangle$ to $\langle A_1, A_2 \rangle$. In this way the set of IR $O$-contracts define a subgraph, $\mathcal{G}_m$ of $\mathcal{H}_m$ with any directed path from $\beta(P)$ to $\beta(Q)$ in $\mathcal{G}_m$ corresponding to a possible IR $O$-contract path from the allocation $\langle P, \mathcal{R} \setminus P \rangle$ to the allocation $\langle Q, \mathcal{R} \setminus Q \rangle$.

## 2.2 $O$-contract Paths − Unrestricted Utility Functions

Our first result clarifies one issue in the presentation of (Sandholm, 1998, Proposition 2): in this an upper bound that is exponential in $m$ is proved on the length of IR $O$-contract paths, i.e. in terms of our notation, (Sandholm, 1998, Proposition 2) establishes an upper bound on $\rho^{\text{max}}(n, m, \Phi, \Psi)$. We now prove a similar order *lower* bound.

**Theorem 3** *Let $\Phi(P, Q)$ be the predicate which holds whenever $\langle P, Q \rangle$ is an IR $O$-contract and $\Psi(P, Q)$ that which holds whenever $\langle P, Q \rangle$ is IR. For $m \geq 7$*

$$\rho^{\text{max}}(2, m, \Phi, \Psi) \ \geq \left( \frac{77}{256} \right) \ 2^m \ - \ 2$$

*Proof.* Consider a path $\mathcal{C} = \langle \alpha_1, \alpha_2, \ldots, \alpha_t \rangle$ in $\mathcal{H}_m$, with the following property[4]

$$\forall\, 1 \leq i < j \leq t\; (j \geq i+2) \;\Rightarrow\; (\alpha_i \text{ and } \alpha_j \text{ differ in at least 2 positions}) \qquad \text{(SC)}$$

e.g. if $m = 4$ then

$$\emptyset, \{r_1\},\; \{r_1, r_3\}, \{r_1, r_2, r_3\}, \{r_2, r_3\}, \{r_2, r_3, r_4\}, \{r_2, r_4\}, \{r_1, r_2, r_4\}$$

is such a path as it corresponds to the sequence $\langle 0000, 1000, 1010, 1110, 0110, 0111, 0101, 1101 \rangle$.

Choose $\mathcal{C}^{(m)}$ to be a *longest* such path that could be formed in $\mathcal{H}_m$, letting $\Delta_m = \langle P^{(1)}, P^{(2)}, \ldots, P^{(t)} \rangle$ be the sequence of allocations with $P^{(i)} = \langle \alpha_i, \overline{\alpha_i} \rangle$. We now define the utility functions $u_1$ and $u_2$ so that for $\gamma \subseteq \mathcal{R}_m$,

$$u_1(\gamma) + u_2(\overline{\gamma}) = \begin{cases} k & \text{if} \quad \gamma = \alpha_k \\ 0 & \text{if} \quad \gamma \notin \{\alpha_1, \alpha_2, \ldots, \alpha_t\} \end{cases}$$

With this choice, the contract path $\Delta_m$ describes the *unique* IR $O$-contract path realising the IR deal $\langle P^{(1)}, P^{(t)} \rangle$: that $\Delta_m$ is an IR $O$-contract path is immediate, since

$$\sigma_u(P^{(i+1)}) = i + 1 > i = \sigma_u(P^{(i)})$$

That it is unique follows from the fact that for all $1 \leq i \leq t$ and $i + 2 \leq j \leq t$, the deal $\langle P^{(i)}, P^{(j)} \rangle$ is not an $O$-contract (hence there are no "short-cuts" possible), and for each $P^{(i)}$ there is exactly one IR $O$-contract that can follow it, i.e. $P^{(i+1)}$.[5]

From the preceding argument it follows that any lower bound on the length of $\mathcal{C}^{(m)}$, i.e. a sequence satisfying the condition (SC), is a lower bound on $\rho^{\max}(2, m, \Phi, \Psi)$. These paths in $\mathcal{H}_m$ were originally studied by Kautz (1958) in the context of coding theory and the lower bound on their length of $(77/256)2^m - 2$ established in (Abbott & Katchalski, 1991). □

**Example 1** *Using the path*

$$\begin{aligned} \mathcal{C}^{(4)} &= \langle 0000, 1000, 1010, 1110, 0110, 0111, 0101, 1101 \rangle \\ &= \langle \alpha_1, \alpha_2, \alpha_3, \alpha_4, \alpha_5, \alpha_6, \alpha_7, \alpha_8 \rangle \end{aligned}$$

*in the resource allocation setting* $\langle \{a_1, a_2\}, \{r_1, r_2, r_3, r_4\}, \langle u_1, u_2 \rangle \rangle$, *if the utility functions are specified as in Table 1 below then* $\sigma_u(\langle \alpha_1, \overline{\alpha_1} \rangle) = 1$ *and* $\sigma_u(\langle \alpha_8, \overline{\alpha_8} \rangle) = 8$. *Furthermore,* $\mathcal{C}^{(4)}$ *describes the unique IR $O$-contract path realising the reallocation* $\langle \langle \alpha_1, \overline{\alpha_1} \rangle, \langle \alpha_8, \overline{\alpha_8} \rangle \rangle$

There are a number of alternative formulations of "rationality" which can also be considered. For example

**Definition 7** *Let* $\delta = \langle P, Q \rangle$ *be a deal.*

---

4. This defines the so-called "snake-in-the-box" codes introduced in (Kautz, 1958).
5. In our example with $m = 4$, the sequence $\langle 0000, 1000, 1001, 1101 \rangle$, although defining an $O$-contract path gives rise to a deal which is not IR, namely that corresponding to $\langle 1000, 1001 \rangle$.

| $S$ | $\mathcal{R}\setminus S$ | $u_1(S)$ | $u_2(\mathcal{R}\setminus S)$ | $\sigma_u$ | | $S$ | $\mathcal{R}\setminus S$ | $u_1(S)$ | $u_2(\mathcal{R}\setminus S)$ | $\sigma_u$ | |
|---|---|---|---|---|---|---|---|---|---|---|---|
| 0000 | 1111 | 1 | 0 | 1 | $\alpha_1$ | 1000 | 0111 | 1 | 1 | 2 | $\alpha_2$ |
| 0001 | 1110 | 0 | 0 | 0 | | 1001 | 0110 | 0 | 0 | 0 | |
| 0010 | 1101 | 0 | 0 | 0 | | 1010 | 0101 | 2 | 1 | 3 | $\alpha_3$ |
| 0011 | 1100 | 0 | 0 | 0 | | 1011 | 0100 | 0 | 0 | 0 | |
| 0100 | 1011 | 0 | 0 | 0 | | 1100 | 0011 | 0 | 0 | 0 | |
| 0101 | 1010 | 4 | 3 | 7 | $\alpha_7$ | 1101 | 0010 | 4 | 4 | 8 | $\alpha_8$ |
| 0110 | 1001 | 3 | 2 | 5 | $\alpha_5$ | 1110 | 0001 | 2 | 2 | 4 | $\alpha_4$ |
| 0111 | 1000 | 3 | 3 | 6 | $\alpha_6$ | 1111 | 0000 | 0 | 0 | 0 | |

Table 1: Utility function definitions for $m = 4$ example.

a. $\delta$ is cooperatively rational *if for every agent, $A_i$, $u_i(Q_i) \geq u_i(P_i)$ and there is at least one agent, $A_j$, for whom $u_j(Q_j) > u_j(P_j)$.*

b. $\delta$ is equitable *if $\min_{i \in \mathcal{A}^\delta} u_i(Q_i) > \min_{i \in \mathcal{A}^\delta} u_i(P_i)$.*

c. $\delta$ is a Pigou-Dalton deal *if $\mathcal{A}^\delta = \{i, j\}$, $u_i(P_i) + u_j(P_j) = u_i(Q_i) + u_j(Q_j)$ and $|u_i(Q_i) - u_j(Q_j)| < |u_i(P_i) - u_j(P_j)|$ (where $|\ldots|$ is absolute value).*

There are a number of views we can take concerning the rationality conditions given in Definition 7. One shared feature is that, unlike the concept of individual rationality for which some provision to compensate agents who suffer a loss in utility is needed, i.e. individual rationality presumes a "money-based" system, the forms defined in Definition 7 allow concepts of "rationality" to be given in "money-free" enviroments. Thus, in a cooperatively rational deal, no agent involved suffers a loss in utility and *at least one* is better off. It may be noted that given the characterisation of Definition 4 it is immediate that any cooperatively rational deal is perforce also individually rational; the converse, however, clearly does not hold in general. In some settings, an equitable deal may be neither cooperatively nor individually rational. One may interpret such deals as one method of reducing inequality between the values agents place on their allocations: for those involved in an equitable deal, it is ensured that the agent who places least value on their current allocation will obtain a resource set which is valued more highly. It may, of course, be the case that some agents suffer a *loss* of utility: the condition for a deal to be equitable limits how great such a loss could be. Finally the concept of Pigou-Dalton deal originates from and has been studied in depth within the theory of exchange economies. This is one of many approaches that have been proposed, again in order to describe deals which reduce inequality between members of an agent society, e.g. (Endriss & Maudet, 2004b). In terms of the definition given, such deals encapsulate the so-called Pigou-Dalton principle in economic theory: that any transfer of income from a wealthy individual to a poorer one should reduce the disparity between them. We note that, in principle, we could define related rationality concepts based on several extensions of this principle that have been suggested, e.g. (Atkinson, 1970; Chateauneaf et al., 2002; Kolm, 1976).

Using the same $O$-contract path constructed in Theorem 3, we need only vary the definitions of the utility functions employed in order to obtain,

**Corollary 1** *For each of the cases below,*

a. $\Phi(\delta)$ *holds if and only if* $\delta$ *is a cooperatively rational O-contract.*
   $\Psi(\delta)$ *holds if and only if* $\delta$ *is cooperatively rational.*

b. $\Phi(\delta)$ *holds if and only if* $\delta$ *is an equitable O-contract.*
   $\Psi(\delta)$ *holds if and only if* $\delta$ *is equitable.*

c. $\Phi(\delta)$ *holds if and only if* $\delta$ *is a Pigou-Dalton O-contract.*
   $\Psi(\delta)$ *holds if and only if* $\delta$ *is a Pigou-Dalton deal.*

$$\rho^{\max}(2, m, \Phi, \Psi) \;\geq\; \left(\frac{77}{256}\right) 2^m \;-\; 2$$

*Proof.* We employ exactly the same sequence of allocations $\Delta_m$ described in the proof of Theorem 3 but modify the utility functions $\langle u_1, u_2 \rangle$ for each case.

a. Choose $\langle u_1, u_2 \rangle$ with $u_2(\gamma) = 0$ for all $\gamma \subseteq \mathcal{R}$ and

$$u_1(\gamma) = \begin{cases} k & \text{if} \quad \gamma = \alpha_k \\ 0 & \text{if} \quad \gamma \notin \{\alpha_1, \ldots, \alpha_t\} \end{cases}$$

The resulting $O$-contract path is cooperatively rational: the utility enjoyed by $A_2$ remains constant while that enjoyed by $A_1$ increases by 1 with each deal. Any deviation from this contract path (employing an alternative $O$-contract) will result in a loss of utility for $A_1$.

b. Choose $\langle u_1, u_2 \rangle$ with $u_2(\gamma) = u_1(\overline{\gamma})$ and

$$u_1(\gamma) = \begin{cases} k & \text{if} \quad \gamma = \alpha_k \\ 0 & \text{if} \quad \gamma \notin \{\alpha_1, \ldots, \alpha_t\} \end{cases}$$

The $O$-contract path is equitable: both $A_1$ and $A_2$ increase their respective utility values by 1 with each deal. Again, any $O$-contract deviating from this will result in both agents losing some utility.

c. Choose $\langle u_1, u_2 \rangle$ as

$$u_1(\gamma) = \begin{cases} k & \text{if} \quad \gamma = \alpha_k \\ 0 & \text{if} \quad \gamma \notin \{\alpha_1, \ldots, \alpha_t\} \end{cases} \quad ; \quad u_2(\gamma) = \begin{cases} 2^m - k & \text{if} \quad \overline{\gamma} = \alpha_k \\ 2^m & \text{if} \quad \overline{\gamma} \notin \{\alpha_1, \ldots, \alpha_t\} \end{cases}$$

To see that the $O$-contract path consists of Pigou-Dalton deals, it suffices to note that $u_1(\alpha_i) + u_2(\overline{\alpha_i}) = 2^m$ for each $1 \leq i \leq t$. In addition, $|u_2(\overline{\alpha_{i+1}}) - u_1(\alpha_{i+1})| = 2^m - 2i - 2$ which is strictly less than $|u_2(\overline{\alpha_i}) - u_1(\alpha_i)| = 2^m - 2i$. Finally, any $O$-contract $\langle P, Q \rangle$ which deviates from this sequence will not be a Pigou-Dalton deal since

$$|u_2(Q_2) - u_1(Q_1)| = 2^m \;>\; |u_2(P_2) - u_1(P_1)|$$

which violates one of the conditions required of Pigou-Dalton deals. □

The construction for two agent settings, easily extends to larger numbers.

**Corollary 2** *For each of the choices of $\langle \Phi, \Psi \rangle$ considered in Theorem 3 and Corollary 1, and all $n \geq 2$,*

$$\rho^{\max}(n, m, \Phi, \Psi) \ \geq \ \left(\frac{77}{256}\right) 2^m \ - \ 2$$

*Proof.* Fix allocations in which $A_1$ is given $\alpha_1$, $A_2$ allocated $\overline{\alpha_1}$, and $A_j$ assigned $\emptyset$ for each $3 \leq j \leq n$. Using identical utility functions $\langle u_1, u_2 \rangle$ as in each of the previous cases, we employ for $u_j$: $u_j(\emptyset) = 1$, $u_j(S) = 0$ whenever $S \neq \emptyset$ ($\langle \Phi, \Psi \rangle$ as in Theorem 3); $u_j(S) = 0$ for all $S$ (Corollary 1(a)); $u_j(\emptyset) = 2^m$, $u_j(S) = 0$ whenever $S \neq \emptyset$ (Corollary 1(b)); and, finally, $u_j(S) = 2^m$ for all $S$, (Corollary 1(c)). Considering a realisation of the $\Psi$-deal $\langle P^{(1)}, P^{(t)} \rangle$ the only $\Phi$-contract path admissible is the path $\Delta_m$ defined in the related proofs. This gives the lower bound stated. □

We note, at this point, some other consequences of Corollary 1 with respect to (Endriss & Maudet, 2004b, Theorems 1, 3), which state

**Fact 1** *We recall that a $\Phi$-path, $\langle P^{(1)}, \ldots, P^{(t)} \rangle$ is* maximal *if for each allocation $Q$, $\langle P^{(t)}, Q \rangle$ is* not *a $\Phi$-deal.*

   a. *If $\langle P^{(1)}, \ldots, P^{(t)} \rangle$ is any maximal path of cooperatively rational deals then $P^{(t)}$ is Pareto optimal.*

   b. *If $\langle P^{(1)}, \ldots, P^{(t)} \rangle$ is any maximal path of equitable deals then $P^{(t)}$ maximises the value $\sigma_e(P) = \min_{1 \leq i \leq n} u_i(P_i)$, i.e. the so-called egalitarian social welfare.*

The sequence of cooperatively rational deals in Corollary 1(a) terminates in the Pareto optimal allocation $P^{(t)}$: the allocation for $A_2$ always has utility 0 and there is no allocation to $A_1$ whose utility can exceed $t$. Similarly, the sequence of equitable deals in Corollary 1(b) terminates in the allocation $P^{(t)}$, for which $\sigma_e(P^{(t)}) = t$ the maximum that can be attained for the instance defined. In both cases, however, the optima are reached by sequences of exponentially many (in $m$) deals: thus, although Fact 1 guarantees convergence of particular deal sequences to optimal states it may be the case, as illustrated in Corollary 1(a–b), that the process of convergence takes considerable time.

## 2.3 *O-contract Paths – Monotone Utility Functions*

We conclude our results concerning $O$-contracts by presenting a lower bound on $\rho^{\max}_{\text{mono}}$, i.e. the length of paths when the utility functions are required to be *monotone*.

In principle one could attempt to construct appropriate monotone utility functions that would have the desired properties with respect to the path used in Theorem 3. It is, however, far from clear whether such a construction is possible. We do not attempt to resolve this question here. Whether an exact translation could be accomplished is, ultimately, a question of purely combinatorial interest: since our aim is to demonstrate that exponential length contract paths are needed with monotone utility functions we are not, primarily, concerned with obtaining an optimal bound.

**Theorem 4** *With $\Phi(P, Q)$ and $\Psi(P, Q)$ be defined as in Theorem 3 and $m \geq 14$*

$$\rho_{\mathrm{mono}}^{\max}(2, m, \Phi, \Psi) \;\geq\; \begin{cases} \left(\frac{77}{128}\right) 2^{m/2} \;-\; 3 & \text{if} \quad m \text{ is even} \\[2mm] \left(\frac{77}{128}\right) 2^{(m-1)/2} \;-\; 3 & \text{if} \quad m \text{ is odd} \end{cases}$$

*Proof.* We describe the details only for the case of $m$ being even: the result when $m$ is odd is obtained by a simple modification which we shall merely provide in outline. Let $m = 2s$ with $s \geq 7$. For any path

$$\Delta_s \;=\; \langle \alpha_1, \alpha_2, \ldots, \alpha_t \rangle$$

in $\mathcal{H}_s$ (where $\alpha_i$ describes a subset of $\mathcal{R}_s$ by an $s$-bit label), the path $double(\Delta_s)$ in $\mathcal{H}_{2s}$ is defined by

$$\begin{aligned} double(\Delta_s) \;&=\; \langle\, \alpha_1\overline{\alpha_1},\; \alpha_2\overline{\alpha_2}\,,\ldots,\; \alpha_i\overline{\alpha_i},\; \alpha_{i+1}\overline{\alpha_{i+1}}\,,\ldots,\; \alpha_t\overline{\alpha_t}\,\rangle \\ &=\; \langle \beta_1, \beta_3, \ldots, \beta_{2i-1}, \beta_{2i+1}, \ldots, \beta_{2t-1} \rangle \end{aligned}$$

(The reason for successive indices of $\beta$ increasing by 2 will become clear subsequently)

Of course, $double(\Delta_s)$ does not describe an $O$-contract path[6]: it is, however, not difficult to interpolate appropriate allocations, $\beta_{2i}$, in order to convert it to such a path. Consider the subsets $\beta_{2i}$ (with $1 \leq i < t$) defined as follows:

$$\beta_{2i} \;=\; \begin{cases} \alpha_{i+1}\overline{\alpha_i} & \text{if} \quad \alpha_i \subset \alpha_{i+1} \\ \alpha_i\overline{\alpha_{i+1}} & \text{if} \quad \alpha_i \supset \alpha_{i+1} \end{cases}$$

If we now consider the path, $ext(\Delta_s)$, within $\mathcal{H}_{2s}$ given by

$$ext(\Delta_s) \;=\; \langle \beta_1, \beta_2, \beta_3\,, \ldots,\; \beta_{2(t-1)}, \beta_{2t-1} \rangle$$

then this satisfies,

    a. If $\Delta_s$ has property (SC) of Theorem 3 in $\mathcal{H}_s$ then $ext(\Delta_s)$ has property (SC) in $\mathcal{H}_{2s}$.

    b. If $j$ is odd then $|\beta_j| = s$.

    c. If $j$ is even then $|\beta_j| = s + 1$.

From (a) and the bounds proved in (Abbott & Katchalski, 1991) we deduce that $ext(\Delta_s)$ can be chosen so that with $P^{(i)}$ denoting the allocation $\langle \beta_i, \overline{\beta_i} \rangle$

    d. $ext(\Delta_s)$ describes an $O$-contract path from $P^{(1)}$ to $P^{(2t-1)}$.

    e. For each pair $\langle i, j \rangle$ with $j \geq i + 2$, the deal $\langle P^{(i)}, P^{(j)} \rangle$ is *not* an $O$-contract.

    f. If $\Delta_s$ is chosen as in the proof of Theorem 3 then the number of deals in $ext(\Delta_s)$ is as given in the statement of the present theorem.

---

6. In terms of the classification described by Sandholm (1998), it contains only *swap* deals ($S$-contracts): each deal swaps exactly one item in $\beta_{2i-1}$ with an item in $\overline{\beta_{2i-1}}$ in order to give $\beta_{2i+1}$.

We therefore fix $\Delta_s$ as the path from Theorem 3 so that in order to complete the proof we need to construct utility functions $\langle u_1, u_2 \rangle$ that are monotone and with which $ext(\Delta_s)$ defines the unique IR $O$-contract path realising the reallocation $\langle P^{(1)}, P^{(2t-1)} \rangle$.

The choice for $u_2$ is relatively simple. Given $S \subseteq \mathcal{R}_{2s}$,

$$u_2(S) \;=\; \left\{ \begin{array}{lll} 0 & \text{if} & |S| \leq s - 2 \\ 2t+1 & \text{if} & |S| = s - 1 \\ 2t+2 & \text{if} & |S| \geq s \end{array} \right.$$

In this $t$ is the number of allocations in $\Delta_s$. The behaviour of $u_2$ is clearly monotone.

The construction for $u_1$ is rather more complicated. Its main idea is to make use of the fact that the size of each set $\beta_i$ occurring in $ext(\Delta_s)$ is very tightly constrained: $|\beta_i|$ is either $s$ or $s + 1$ according to whether $i$ is odd or even. We first demonstrate that each set of size $s + 1$ can have at most two strict subsets (of size $s$) occurring within $ext(\Delta_s)$: thus, every $S$ of size $s + 1$ has exactly 2 or 1 or 0 subsets of size $s$ on $ext(\Delta_s)$. To see this suppose the contrary. Let $\gamma$, $\beta_{2i-1}$, $\beta_{2j-1}$, and $\beta_{2k-1}$ be such that $|\gamma| = s + 1$ with

$$\beta_{2i-1} \subset \gamma \;\; ; \;\; \beta_{2j-1} \subset \gamma \;\; ; \;\; \beta_{2k-1} \subset \gamma$$

Noting that $\beta_{2i-1} = \alpha_i \overline{\alpha_i}$ and that $\Delta_s$ has the property (SC) it must be the case that (at least) two of the $s$-bit labels from $\{\alpha_i, \alpha_j, \alpha_k\}$ differ in at least two positions. Without loss of generality suppose this is true of $\alpha_i$ and $\alpha_k$. As a result we deduce that the sets $\beta_{2i-1}$ and $\beta_{2k-1}$ have at most $s-2$ elements in common, i.e. $|\beta_{2i-1} \cap \beta_{2k-1}| \leq s-2$: $\beta_{2i-1} = \alpha_i \overline{\alpha_i}$ and $\beta_{2k-1} = \alpha_k \overline{\alpha_k}$ so in any position at which $\alpha_i$ differs from $\alpha_k$, $\overline{\alpha_i}$ differs from $\overline{\alpha_k}$ at exactly the same position. In total $|\beta_{2i-1} \setminus \beta_{2k-1}| \geq 2$, i.e. there are (at least) two elements of $\beta_{2i-1}$ that do not occur in $\beta_{2k-1}$; and in the same way $|\beta_{2k-1} \setminus \beta_{2i-1}| \geq 2$, i.e. there are (at least) two elements of $\beta_{2k-1}$ that do not occur in $\beta_{2i-1}$. The set $\gamma$, however, has only $s + 1$ members and so cannot have *both* $\beta_{2i-1}$ and $\beta_{2k-1}$ as subsets: this would require

$$\beta_{2i-1} \cap \beta_{2k-1} \;\cup\; \beta_{2i-1} \setminus \beta_{2k-1} \;\cup\; \beta_{2k-1} \setminus \beta_{2i-1} \subseteq \gamma$$

but, as we have just seen,

$$|\; \beta_{2i-1} \cap \beta_{2k-1} \;\cup\; \beta_{2i-1} \setminus \beta_{2k-1} \;\cup\; \beta_{2k-1} \setminus \beta_{2i-1} \;| \;\; \geq \;\; s+2$$

One immediate consequence of the argument just given is that for any set $\gamma$ of size $s+1$ there are exactly two strict subsets of $\gamma$ occurring on $ext(\Delta_s)$ if and only if $\gamma = \beta_{2i-1} \cup \beta_{2i+1} = \beta_{2i}$ for some value of $i$ with $1 \leq i < t$. We can now characterise each subset of $\mathcal{R}_{2s}$ of size $s + 1$ as falling into one of three categories.

C1. *Good* sets, given by $\{\gamma \;:\; \gamma = \beta_{2i}\}$.

C2. *Digressions*, consisting of

$$\{\; \gamma \;:\; \beta_{2i-1} \subset \gamma, \gamma \neq \beta_{2i} \text{ and } i < t \}$$

C3. *Inaccessible* sets, consisting of

$$\{\; \gamma \;:\; \gamma \text{ is neither } \textit{Good} \text{ nor a } \textit{Digression} \}$$

*Good* sets are those describing allocations to $A_1$ within the path defined by $ext(\Delta_s)$; *Digressions* are the allocations that could be reached using an $O$-contract from a set of size $s$ on $ext(\Delta_s)$, i.e. $\beta_{2i-1}$, but differ from the set that actually occurs in $ext(\Delta_s)$, i.e. $\beta_{2i}$. Finally, *Inaccessible* sets are those that do not occur on $ext(\Delta_s)$ *and* cannot be reached via an $O$-contract from any set on $ext(\Delta_s)$. We note that we view any set of size $s+1$ that *could* be reached by an $O$-contract from $\beta_{2t-1}$ as being inaccessible: in principle it is possible to extend the $O$-contract path beyond $\beta_{2t-1}$, however, we choose not complicate the construction in this way.

We now define $u_1$ as

$$u_1(\gamma) \;=\; \begin{cases} 2i-1 & \text{if} \quad \gamma = \beta_{2i-1} \\ 2i+1 & \text{if} \quad \gamma = \beta_{2i} \\ 2i & \text{if} \quad |\gamma| = s+1 \text{ and } \gamma \text{ is a } \textit{Digression} \text{ from } \beta_{2i-1} \\ 0 & \text{if} \quad |\gamma| \leq s-1 \\ 0 & \text{if} \quad |\gamma| = s \text{ and } \gamma \notin ext(\Delta_s) \\ 2t-1 & \text{if} \quad \gamma \text{ is } \textit{Inaccessible} \text{ or } |\gamma| \geq s+2 \end{cases}$$

It remains only to prove for these choices of $\langle u_1, u_2 \rangle$ that the $O$-contract path $\langle P^{(1)}, \ldots, P^{(2t-1)} \rangle$ defined from $ext(\Delta_s)$ is the unique IR $O$-contract path realising the IR deal $\langle P^{(1)}, P^{(2t-1)} \rangle$ and that $u_1$ is monotone.

To show that $\langle P^{(1)}, \ldots, P^{(2t-1)} \rangle$ is IR we need to demonstrate

$$\forall\; 1 \leq j < 2t-1 \quad u_1(\beta_j) + u_2(\overline{\beta_j}) \;<\; u_1(\beta_{j+1}) + u_2(\overline{\beta_{j+1}})$$

We have via the definition of $\langle u_1, u_2 \rangle$

$$\begin{aligned} u_1(\beta_{2i-1}) + u_2(\overline{\beta_{2i-1}}) &= 2(t+i)+1 \\ &< u_1(\beta_{2i}) + u_2(\overline{\beta_{2i}}) \\ &= 2(t+i)+2 \\ &< u_1(\beta_{2i+1}) + u_2(\overline{\beta_{2i+1}}) \\ &= 2(t+i)+3 \end{aligned}$$

Thus, via Definition 4, it follows that $ext(\Delta_s)$ gives rise to an IR $O$-contract path.

To see that this path is the unique IR $O$-contract path implementing $\langle P^{(1)}, P^{(2t-1)} \rangle$, consider any position $P^{(j)} = \langle \beta_j, \overline{\beta_j} \rangle$ and allocation $Q$ other than $P^{(j+1)}$ or $P^{(j-1)}$. It may be assumed that the deal $\langle P^{(j)}, Q \rangle$ is an $O$-contract. If $j = 2i-1$ then $\sigma_u(P^{(2i-1)}) = 2(t+i)+1$ and $|\beta_j| = s$. Hence $|Q_1| \in \{s-1, s+1\}$. In the former case, $u_1(Q_1) = 0$ and $u_2(Q_2) = 2t+2$ from which $\sigma_u(Q) = 2t+2$ and thus $\langle P^{(j)}, Q \rangle$ is not IR. In the latter case $u_1(Q_1) = 2i$ since $Q_1$ is a *Digression* from $\beta_{2i-1}$ and $u_2(Q_2) = 2t+1$ giving $\sigma_u(Q) = 2(t+i)+1$. Again $\langle P^{(j)}, Q \rangle$ fails to be IR since $Q$ fails to give any increase in the value of $\sigma_u$. We are left with the case $j = 2i$ so that $\sigma_u(P^{(2i)}) = 2(t+i)+2$ and $|\beta_j| = s+1$. Since $\langle P^{(j)}, Q \rangle$ is assumed to be an $O$-contract this gives $|Q_1| \in \{s, s+2\}$. For the first possibility $Q_1$ could not be a set on $ext(\Delta_s)$: $\beta_{2i-1}$ and $\beta_{2i+1}$ are both subsets of $\beta_{2i}$ and there can be at most two such subsets occurring on $ext(\Delta_s)$. It follows, therefore, that $u_1(Q_1) = 0$ giving $\sigma_u(Q) = 2t+2$ so that $\langle P^{(j)}, Q \rangle$ is not IR. In the second possibility, $u_1(Q_1) = 2t-1$ but $u_2(Q_2) = 0$ as $|Q_2| = s-2$ so the deal would result in an overall loss. We deduce that for each $P^{(j)}$ the only IR $O$-contract consistent with it is the deal $\langle P^{(j)}, P^{(j+1)} \rangle$.

The final stage is to prove that the utility function $u_1$ is indeed a *monotone* function. Suppose $S$ and $T$ are subsets of $\mathcal{R}_{2s}$ with $S \subset T$. We need to show that $u_1(S) \leq u_1(T)$. We may assume that $|S| = s$, that $S$ occurs as some set within $ext(\Delta_s)$, and that $|T| = s + 1$. If $|S| < s$ or $|S| = s$ but does not occur on $ext(\Delta_s)$ we have $u_1(S) = 0$ and the required inequality holds; if $|S| \geq s + 1$ then in order for $S \subset T$ to be possible we would need $|T| \geq s + 2$, which would give $u_1(T) = 2t - 1$ and this is the maximum value that any subset is assigned by $u_1$. We are left with only $|S| = s$, $|T| = s + 1$ and $S$ on $ext(\Delta_s)$ to consider. It has already been shown that there are at most two subsets of $T$ that can occur on $ext(\Delta_s)$. Consider the different possibilities:

a. $T = \beta_{2i}$ so that exactly two subsets of $T$ occur in $ext(\Delta_s)$: $\beta_{2i-1}$ and $\beta_{2i+1}$. Since $u_1(\beta_{2i}) = 2i + 1$ and this is at least $\max\{u_1(\beta_{2i-1}), u_1(\beta_{2i+1})\}$, should $S$ be either of $\beta_{2i-1}$ or $\beta_{2i+1}$ then $u_1(S) \leq u_1(T)$ as required.

b. $T$ is a *Digression* from $S = \beta_{2i-1}$, so that $u_1(T) = 2i$ and $u_1(S) = 2i - 1$ and, again, $u_1(S) \leq u_1(T)$.

We deduce that $u_1$ is monotone completing our lower bound proof for $\rho_{\mathrm{mono}}^{\max}$ for even values of $m$.

We conclude by observing that a similar construction can be used if $m = 2s + 1$ is odd: use the path $ext(\Delta_s)$ described above but modifying it so that one resource ($r_m$) is always held by $A_2$. Only minor modifications to the utility function definitions are needed. □

**Example 2** *For $s = 3$, we can choose $\Delta_3 = \langle 000, 001, 101, 111, 110 \rangle$ so that $t = 5$. This gives double$(\Delta_3)$ as*

$$\langle 000111, 001110, 101010, 111000, 110001 \rangle$$

*with the O-contract path being defined from $ext(\Delta_3)$ which is*

$$\langle 000111, 001111, 001110, 101110, 101010, 111010, 111000, 111001, 110001 \rangle$$
$$= \langle \beta_1, \beta_2, \beta_3, \beta_4, \beta_5, \beta_6, \beta_7, \beta_8, \beta_9 \rangle$$

*Considering the 15 subsets of size $s + 1 = 4$, gives*

| | | |
|---|---|---|
| *Good* | = | {001111, 101110, 111010, 111001} |
| *Digression* | = | {010111, 100111, 101011, 011110, 111100} |
| *Inaccessible* | = | {011011, 011101, 101101, 110110, 110011, 110101} |

*Notice that both of the sets in {110011, 110101} are Inaccessible: in principle we could continue from $\beta_9 = 110001$ using either, however, in order to simplify the construction the path is halted at $\beta_9$.*

*Following the construction presented in Theorem 4, gives the following utility function definitions with $S \subseteq \mathcal{R} = \{r_1, r_2, r_3, r_4, r_5, r_6\}$.*

$$u_2(S) = \begin{cases} 0 & \text{if} \quad |S| \leq 1 \\ 11 & \text{if} \quad |S| = 2 \\ 12 & \text{if} \quad |S| \geq 3 \end{cases}$$

*For $u_1$ we obtain*

$$
u_1(S) \;=\; \left\{
\begin{array}{lll}
0 & \text{if} & |S| \le 2 \\
0 & \text{if} & |S| = 3 \text{ and } S \notin \{000111, 001110, 101010, 111000, 110001\} \\
1 & \text{if} & S = 000111 \;\; (\beta_1) \\
2 & \text{if} & S = 010111 \;\; (\text{digression from } \beta_1) \\
2 & \text{if} & S = 100111 \;\; (\text{digression from } \beta_1) \\
3 & \text{if} & S = 001111 \;\; (\beta_2) \\
3 & \text{if} & S = 001110 \;\; (\beta_3) \\
4 & \text{if} & S = 011110 \;\; (\text{digression from } \beta_3) \\
5 & \text{if} & S = 101110 \;\; (\beta_4) \\
5 & \text{if} & S = 101010 \;\; (\beta_5) \\
6 & \text{if} & S = 101011 \;\; (\text{digression from } \beta_5) \\
7 & \text{if} & S = 111010 \;\; (\beta_6) \\
7 & \text{if} & S = 111000 \;\; (\beta_7) \\
8 & \text{if} & S = 111100 \;\; (\text{digression from } \beta_7) \\
9 & \text{if} & S = 111001 \;\; (\beta_8) \\
9 & \text{if} & S = 110001 \;\; (\beta_9) \\
9 & \text{if} & |S| \ge 5 \text{ or } S \in \{011011, 011101, 101101, 110110, 110011, 110101\}
\end{array}
\right.
$$

The monotone utility functions, $\langle u_1, u_2 \rangle$, employed in proving Theorem 4 are defined so that the path arising from $ext(\Delta_s)$ is IR: in the event of either agent suffering a loss of utility the gain made by the other is sufficient to provide a compensatory payment. A natural question that now arises is whether the bound obtained in Theorem 4 can be shown to apply when the rationality conditions preclude any monetary payment, e.g. for cases where the concept of rationality is one of those given in Definition 7. Our next result shows that if we set the rationality condition to enforce cooperatively rational or equitable deals then the bound of Theorem 4 still holds.

**Theorem 5** *For each of the cases below and $m \ge 14$*

    *a. $\Phi(\delta)$ holds if and only if $\delta$ is a cooperatively rational O-contract.*
       *$\Psi(\delta)$ holds if and only if $\delta$ is cooperatively rational.*

    *b. $\Phi(\delta)$ holds if and only if $\delta$ is an equitable O-contract.*
       *$\Psi(\delta)$ holds if and only if $\delta$ is equitable.*

$$
\rho_{\text{mono}}^{\max}(2, m, \Phi, \Psi) \;\ge\; \left\{
\begin{array}{lll}
\left(\frac{77}{128}\right) 2^{m/2} \;-\; 3 & \text{if} & m \text{ is even} \\[2mm]
\left(\frac{77}{128}\right) 2^{(m-1)/2} \;-\; 3 & \text{if} & m \text{ is odd}
\end{array}
\right.
$$

*Proof.* We again illustrate the constructions only for the case of $m$ being even, noting the modification to deal with odd values of $m$ outlined at the end of the proof of Theorem 4. The path $ext(\Delta_s)$ is used for both cases.

For (a), we require $\langle u_1, u_2 \rangle$ to be defined as monotone functions with which $ext(\Delta_s)$ will be the unique cooperatively rational $O$-contract path to realise the cooperatively rational deal $\langle P^{(1)}, P^{(2t-1)} \rangle$ where $P^{(j)} = \langle \beta_j, \overline{\beta_j} \rangle$. In this case we set $\langle u_1, u_2 \rangle$ to be,

$$\langle u_1(\gamma), u_2(\overline{\gamma}) \rangle = \begin{cases} \langle i, i \rangle & \text{if} \quad \gamma = \beta_{2i-1} \\ \langle i+1, i \rangle & \text{if} \quad \gamma = \beta_{2i} \\ \langle i, i-1 \rangle & \text{if} \quad |\gamma| = s+1 \text{ and } \gamma \text{ is a } Digression \text{ from } \beta_{2i-1} \\ \langle 0, 2t-1 \rangle & \text{if} \quad |\gamma| \leq s-1 \\ \langle 0, 2t-1 \rangle & \text{if} \quad |\gamma| = s \text{ and } \gamma \notin ext(\Delta_s) \\ \langle 2t-1, 0 \rangle & \text{if} \quad \gamma \text{ is } Inaccessible \text{ or } |\gamma| \geq s+2 \end{cases}$$

Since,

$$\begin{array}{rcl} \langle u_1(\beta_{2i-1}), u_2(\overline{\beta_{2i-1}}) \rangle & = & \langle i, i \rangle \\ \langle u_1(\beta_{2i}), u_2(\overline{\beta_{2i}}) \rangle & = & \langle i+1, i \rangle \\ \langle u_1(\beta_{2i+1}), u_2(\overline{\beta_{2i+1}}) \rangle & = & \langle i+1, i+1 \rangle \end{array}$$

it is certainly the case that $\langle P^{(1)}, P^{(2t-1)} \rangle$ and all deals on the $O$-contract path defined by $ext(\Delta_s)$ are cooperatively rational. Furthermore if $Q = \langle \gamma, \overline{\gamma} \rangle$ is any allocation other than $P^{(j+1)}$ then the deal $\langle P^{(j)}, Q \rangle$ will fail to be a cooperatively rational $O$-contract. For suppose the contrary letting $\langle P^{(j)}, Q \rangle$ without loss of generality be an $O$-contract, with $Q \notin \{P^{(j-1)}, P^{(j+1)}\}$ – we can rule out the former case since we have already shown such an deal is not cooperatively rational. If $j = 2i-1$ so that $\langle u_1(\beta_j), u_2(\overline{\beta_j}) \rangle = \langle i, i \rangle$ then $|\gamma| \in \{s-1, s+1\}$: the former case leads to a loss in utility for $A_1$, the latter, (since $\gamma$ is a $Digression$ from $\beta_{2i-1}$) a loss in utility for $A_2$. Similarly, if $j = 2i$ so that $\langle u_1(\beta_j), u_2(\overline{\beta_j}) \rangle = \langle i+1, i \rangle$ then $|\gamma| \in \{s, s+2\}$: for the first $\gamma \notin ext(\Delta_s)$ leading to a loss of utility for $A_1$; the second results in a loss of utility for $A_2$. It follows that the path defined by $ext(\Delta_s)$ is the unique cooperatively rational $O$-contract path that realises $\langle P^{(1)}, P^{(2t-1)} \rangle$.

It remains only to show that these choices for $\langle u_1, u_2 \rangle$ define monotone utility functions.

Consider $u_1$ and suppose $S$ and $T$ are subsets of $\mathcal{R}_{2s}$ with $S \subset T$. If $|S| \leq s-1$, or $S$ does not occur on $ext(\Delta_s)$ then $u_1(S) = 0$. If $|T| \geq s+2$ or is $Inaccessible$ then $u_1(T) = 2t-1$ which is the maximum value attainable by $u_1$. So we may assume that $|S| = s$, occurs on $ext(\Delta_s)$, i.e. $S = \beta_{2i-1}$, for some $i$, and that $|T| = s+1$ and is either a $Good$ set or a $Digression$. From the definition of $u_1$, $u_1(S) = i$: if $T \in \{\beta_{2i}, \beta_{2i-2}\}$ then $u_1(T) \geq i = u_1(S)$; if $T$ is a $Digression$ from $\beta_{2i-1}$ then $u_1(T) = i = u_1(S)$. We deduce that if $S \subseteq T$ then $u_1(S) \leq u_1(T)$, i.e. the utility function is monotone.

Now consider $u_2$ with $S$ and $T$ subsets of $\mathcal{R}_{2s}$ having $S \subset T$. If $|T| \geq s+1$ or $\mathcal{R}_{2s} \setminus T$ does not occur in $ext(\Delta_s)$ then $u_2(T) = 2t-1$ its maximal value. If $|S| \leq s-2$ or $\mathcal{R}_{2s} \setminus S$ is $Inaccessible$ then $u_2(S) = 0$. Thus we may assume that $T = \overline{\beta_{2i-1}}$ giving $u_2(T) = i$ and $|S| = s-1$, so that $\mathcal{R}_{2s} \setminus S$ is either a $Digression$ or one of the $Good$ sets $\{\beta_{2i}, \beta_{2i-2}\}$. If $\mathcal{R}_{2s} \setminus S$ is a $Digression$ then $u_2(S) = i-1$; if it is the $Good$ set $\beta_{2i-2}$ then $u_2(S) = i-1 < u_2(T)$; if it is the $Good$ set $\beta_{2i}$ then $u_2(S) = i = u_2(T)$. It follows that $u_2$ is monotone completing the proof of part (a).

For (b) we use,

$$\langle u_1(\gamma), u_2(\overline{\gamma})\rangle \;=\; \begin{cases} \langle 2i-1, 2i\rangle & \text{if} \quad \gamma = \beta_{2i-1} \\ \langle 2i+1, 2i\rangle & \text{if} \quad \gamma = \beta_{2i} \\ \langle 2i, 2i-1\rangle & \text{if} \quad |\gamma| = s+1 \text{ and } \gamma \text{ is a } Digression \text{ from } \beta_{2i-1} \\ \langle 0, 2t-1\rangle & \text{if} \quad |\gamma| \leq s-1 \\ \langle 0, 2t-1\rangle & \text{if} \quad |\gamma| = s \text{ and } \gamma \notin ext(\Delta_s) \\ \langle 2t-1, 0\rangle & \text{if} \quad \gamma \text{ is } Inaccessible \text{ or } |\gamma| \geq s+2 \end{cases}$$

These choices give $ext(\Delta_s)$ as the unique equitable $O$-contract path to realise the equitable deal $\langle P^{(1)}, P^{(2t-1)}\rangle$, since

$$\begin{aligned} \min\{u_1(\beta_{2i-1}), u_2(\overline{\beta_{2i-1}})\} &= 2i-1 \\ \min\{u_1(\beta_{2i}), u_2(\overline{\beta_{2i}})\} &= 2i \\ \min\{u_1(\beta_{2i+1}), u_2(\overline{\beta_{2i+1}})\} &= 2i+1 \end{aligned}$$

each deal $\langle P^{(j)}, P^{(j+1)}\rangle$ is equitable. If $Q = \langle \gamma, \overline{\gamma}\rangle$ is any allocation other than $P^{(j+1)}$ then the deal $\langle P^{(j)}, Q\rangle$ is not an equitable $O$-contract. Assume that $\langle P^{(j)}, Q\rangle$ is an $O$-contract, and that $Q \notin \{P^{(j-1)}, P^{(j+1)}\}$. If $j = 2i-1$, so that $P^{(j)} = \langle \beta_{2i-1}, \overline{\beta_{2i-1}}\rangle$ and $\min\{u_1(\beta_{2i-1}), u_2(\overline{\beta_{2i-1}})\} = 2i-1$ then $|\gamma| \in \{s-1, s+1\}$. In the first of these $\min\{u_1(\gamma), u_2(\overline{\gamma})\} = 0$; in the second $\min\{u_1(\gamma), u_2(\overline{\gamma})\} = 2i-1$ since $\gamma$ must be a $Digression$. This leaves only $j = 2i$ with $P^{(j)} = \langle \beta_{2i}, \overline{\beta_{2i}}\rangle$ and $\min\{u_1(\beta_{2i}), u_2(\overline{\beta_{2i}})\} = 2i$. For this, $|\gamma| \in \{s, s+2\}$: if $|\gamma| = s$ then $\min\{u_1(\gamma), u_2(\overline{\gamma})\} \leq 2i-1$ (with equality when $\gamma = \beta_{2i-1}$); if $|\gamma| = s+2$ then $\min\{u_1(\gamma), u_2(\overline{\gamma})\} = 0$. In total these establish that $ext(\Delta_s)$ is the unique equitable $O$-contract path realising the equitable deal $\langle P^{(1)}, P^{(2t-1)}\rangle$.

That the choices for $\langle u_1, u_2\rangle$ describe monotone utility functions can be shown by a similar argument to that of part (a). □

**Example 3** *For $s = 3$ using the same $O$-contract path $ext(\Delta_3)$ as the previous example, i.e.*

$$\begin{aligned} &\langle 000111, 001111, 001110, 101110, 101010, 111010, 111000, 111001, 110001\rangle \\ =\; &\langle \beta_1, \beta_2, \beta_3, \beta_4, \beta_5, \beta_6, \beta_7, \beta_8, \beta_9\rangle \end{aligned}$$

*For $\langle u_1, u_2\rangle$ in (a) we obtain*

$$\langle u_1(S), u_2(\mathcal{R}\setminus S)\rangle = \begin{cases} \langle 0, 9\rangle & \text{if} \quad |S| \leq 2 \\ \langle 0, 9\rangle & \text{if} \quad |S| = 3 \text{ and } S \notin \{000111, 001110, 101010, 111000, 110001\} \\ \langle 1, 1\rangle & \text{if} \quad S = 000111 \quad (\beta_1) \\ \langle 1, 0\rangle & \text{if} \quad S = 010111 \text{ digression from } \beta_1 \\ \langle 1, 0\rangle & \text{if} \quad S = 100111 \text{ digression from } \beta_1 \\ \langle 2, 1\rangle & \text{if} \quad S = 001111 \quad (\beta_2) \\ \langle 2, 2\rangle & \text{if} \quad S = 001110 \quad (\beta_3) \\ \langle 2, 1\rangle & \text{if} \quad S = 011110 \text{ digression from } \beta_3 \\ \langle 3, 2\rangle & \text{if} \quad S = 101110 \quad (\beta_4) \\ \langle 3, 3\rangle & \text{if} \quad S = 101010 \quad (\beta_5) \\ \langle 3, 2\rangle & \text{if} \quad S = 101011 \text{ digression from } \beta_5 \\ \langle 4, 3\rangle & \text{if} \quad S = 111010 \quad (\beta_6) \\ \langle 4, 4\rangle & \text{if} \quad S = 111000 \quad (\beta_7) \\ \langle 4, 3\rangle & \text{if} \quad S = 111100 \text{ digression from } \beta_7 \\ \langle 5, 4\rangle & \text{if} \quad S = 111001 \quad (\beta_8) \\ \langle 5, 5\rangle & \text{if} \quad S = 110001 \quad (\beta_9) \\ \langle 9, 0\rangle & \text{if} \quad |S| \geq 5 \text{ or } S \in \{011011, 011101, 101101, 110110, 110011, 110101\} \end{cases}$$

*Similarly, in (b)*

$$\langle u_1(S), u_2(\mathcal{R} \setminus S) \rangle = \begin{cases} \langle 0, 9 \rangle & \text{if} & |S| \le 2 \\ \langle 0, 9 \rangle & \text{if} & |S| = 3 \text{ and } S \notin \{000111, 001110, 101010, 111000, 110001\} \\ \langle 1, 2 \rangle & \text{if} & S = 000111 \quad (\beta_1) \\ \langle 2, 1 \rangle & \text{if} & S = 010111 \text{ digression from } \beta_1 \\ \langle 2, 1 \rangle & \text{if} & S = 100111 \text{ digression from } \beta_1 \\ \langle 3, 2 \rangle & \text{if} & S = 001111 \quad (\beta_2) \\ \langle 3, 4 \rangle & \text{if} & S = 001110 \quad (\beta_3) \\ \langle 4, 3 \rangle & \text{if} & S = 011110 \text{ digression from } \beta_3 \\ \langle 5, 4 \rangle & \text{if} & S = 101110 \quad (\beta_4) \\ \langle 5, 6 \rangle & \text{if} & S = 101010 \quad (\beta_5) \\ \langle 6, 5 \rangle & \text{if} & S = 101011 \text{ digression from } \beta_5 \\ \langle 7, 6 \rangle & \text{if} & S = 111010 \quad (\beta_6) \\ \langle 7, 8 \rangle & \text{if} & S = 111000 \quad (\beta_7) \\ \langle 8, 7 \rangle & \text{if} & S = 111100 \text{ digression from } \beta_7 \\ \langle 9, 8 \rangle & \text{if} & S = 111001 \quad (\beta_8) \\ \langle 9, 10 \rangle & \text{if} & S = 110001 \quad (\beta_9) \\ \langle 9, 0 \rangle & \text{if} & |S| \ge 5 \text{ or } S \in \{011011, 011101, 101101, 110110, 110011, 110101\} \end{cases}$$

That we can demonstrate similar extremal behaviours for contract path length with rationality constraints in both money-based (individual rationality) and money-free (cooperative rationality, equitable) settings irrespective of whether monotonicity properties are assumed, has some interesting parallels with other contexts in which monotonicity is relevant. In particular we can observe that in common with the complexity results already noted from (Dunne et al., 2003) – deciding if an allocation is Pareto optimal, if an allocation maximises $\sigma_u$, or if an IR $O$-contract path exists – requiring utility functions to be monotone does not result in a setting which is computationally more tractable.

## 3. $M(k)$-contract paths

We now turn to similar issues with respect to $M(k)$-contracts, recalling that in one respect these offer a form of deal that does not fit into the classification of Sandholm (1998). This classification defines four forms of contract type: $O$-contracts, as considered in the previous section; $S$-contracts, that involve exactly 2 agents swapping single resources; $C$-contracts, in which one agent tranfers *at least* two of its resources to another; and $M$-contracts in which *three or more* agents reallocate their resource holding amongst themselves. Our definition of $M(k)$-contracts permits *two* agents to exchange resources (thus are not $M$-contracts in Sandholm's (1998) scheme) and the deals permitted are not restricted to $O$, $S$, and $C$-contracts. In one regard, however, $M(k)$-contracts are not as general as $M$-contracts since a preset bound $(k)$ is specified for the number of agents involved.

Our main result on $M(k)$-contract paths is the following development of Theorem 3.

**Theorem 6** *Let $\Phi_k(P, Q)$ be the predicate which holds whenever $\langle P, Q \rangle$ is an IR $M(k)$-contract. For all $k \ge 3$, $n \ge k$ and $m \ge \binom{k}{2}$, there is a resource allocation setting $\langle \mathcal{A}, \mathcal{R}, \mathcal{U} \rangle$ and an IR deal $\delta = \langle P, Q \rangle$ for which,*

$$\begin{aligned} L^{\text{opt}}(\delta, \langle \mathcal{A}, \mathcal{R}, \mathcal{U} \rangle, \Phi_k) &= 1 & (a) \\ L^{\text{opt}}(\delta, \langle \mathcal{A}, \mathcal{R}, \mathcal{U} \rangle, \Phi_{k-1}) &\ge 2^{\lfloor 2m/k(k-1) \rfloor} - 1 & (b) \\ L^{\text{opt}}(\delta, \langle \mathcal{A}, \mathcal{R}, \mathcal{U} \rangle, \Phi_{k-2}) & \quad \text{is undefined} & (c) \end{aligned}$$

Before presenting the proof, we comment about the formulation of the theorem statement and give an overview of the proof structure.

We first note that the lower bounds (where defined) have been phrased in terms of the function $L^{\text{opt}}$ as opposed to $\rho^{\max}$ used in the various results on $O$-contract paths in Section 2.2. It is, of course, the case that the bound claimed for $L^{\text{opt}}(\delta, \langle \mathcal{A}, \mathcal{R}, \mathcal{U} \rangle, \Phi_{k-1})$ will also be a lower bound on $\rho^{\max}(n, m, \Phi_{k-1}, \Psi)$ when $n \geq k$ and $\Psi(P, Q)$ holds whenever the deal $\langle P, Q \rangle$ is IR. The statement of Theorem 6, however, claims rather more than this, namely that a *specific* resource allocation setting $\langle \mathcal{A}, \mathcal{R}, \mathcal{U} \rangle$ can be defined for each $n \geq k$ and each $m$, together with an IR deal $\langle P, Q \rangle$ in such a way that: $\langle P, Q \rangle$ can be achieved by a *single* $M(k)$-contract and *cannot* be realised by an IR $M(k-2)$-contract path. Recalling that $L^{\text{opt}}$ is a *partial* function, the latter property is equivalent to the claim made in part (c) for the deal $\langle P, Q \rangle$ of the theorem statement. Furthermore, this same deal although achievable by an IR $M(k-1)$-contract path can be so realised only by one whose length is as given in part (b) of the theorem statement.

Regarding the proof itself, there are a number of notational complexities which we have attempted to ameliorate by making some simplifying assumptions concerning the relationship between $m$ – the size of the resource set $\mathcal{R}$ – and $k$ – the number of agents which are needed to realise $\langle P, Q \rangle$ in a *single* IR deal. In particular, we shall assume that $m$ is an exact multiple of $\binom{k}{2}$. We observe that by employing a similar device to that used in the proof of Theorem 4 we can deal with cases for which $m$ does not have this property: if $m = s \binom{k}{2} + q$ for integer values $s \geq 1$ and $1 \leq q < \binom{k}{2}$, we simply employ exactly the same construction using $m - q$ resources with the "missing" $q$ resources from $\mathcal{R}_m$ being allocated to $A_1$ and never being reallocated within the $M(k-1)$-contract path. This approach accounts for the rounding operation ($\lfloor \ldots \rfloor$) in the exponent term of the lower bound. We shall also assume that the number of agents in $\mathcal{A}$ is exactly $k$. Within the proof we use a running example for which $k = 4$ and $m = 18 = 3 \times 6$ to illustrate specific features.

We first give an outline of its structure.

Given $\langle \mathcal{A}, \mathcal{R}, \mathcal{U} \rangle$ a resource allocation setting involving $k$ agents and $m$ resources, our aim is to define an IR $M(k-1)$-contract path

$$\Delta = \langle P^{(1)}, P^{(2)}, \ldots, P^{(t)} \rangle$$

that realises the IR $M(k)$ deal $\langle P^{(1)}, P^{(t)} \rangle$. We will use $d$ to index particular allocations within $\Delta$, so that $1 \leq d \leq t$.

In order to simplify the presentation we employ a setting in which the $k$ agents are $\mathcal{A} = \{A_0, A_1, \ldots, A_{k-1}\}$. Recalling that $m = s \binom{k}{2}$, the resource set $\mathcal{R}_m$ is formed by the union of $\binom{k}{2}$ pairwise disjoint sets of size $s$. Given distinct values $i$ and $j$ with $0 \leq i < j \leq k-1$, we use $\mathcal{R}^{i,j}$ to denote one of these subsets with $\{r_1^{\{i,j\}}, r_2^{\{i,j\}}, \ldots, r_s^{\{i,j\}}\}$ the $s$ resources that form $\mathcal{R}^{\{i,j\}}$.

There are two main ideas underpinning the structure of each $M(k-1)$-contract in $\Delta$.

Firstly, in the initial and subsequent allocations, the resource set $\mathcal{R}^{\{i,j\}}$ is partitioned between $A_i$ and $A_j$ and any reallocation of resources between $A_i$ and $A_j$ that takes place within the deal $\langle P^{(d)}, P^{(d+1)} \rangle$ will involve *only* resources in this set. Thus, for every allocation $P^{(d)}$ and each pair $\{i, j\}$, if $h \notin \{i, j\}$ then $P_h^{(d)} \cap \mathcal{R}^{\{i,j\}} = \emptyset$. Furthermore, for $\delta = \langle P^{(d)}, P^{(d+1)} \rangle$ should *both* $A_i$ and $A_j$ be involved, i.e. $\{A_i, A_j\} \subseteq \mathcal{A}^\delta$, then this real-location of $\mathcal{R}^{\{i,j\}}$ between $A_i$ and $A_j$ will be an $O$-contract. That is, either exactly one element of $\mathcal{R}^{\{i,j\}}$ will be moved from $P_i^{(d)}$ to become a member of the allocation $P_j^{(d+1)}$ or exactly one element of $\mathcal{R}^{\{i,j\}}$ will be moved from $P_j^{(d)}$ to become a member of the allocation $P_i^{(d+1)}$. In total, every $M(k-1)$-contract $\delta$ in $\Delta$ consists of a *simultaneous* implementation of $\binom{k-1}{2}$ $O$-contracts: a single $O$-contract for each of the distinct pairs $\{A_i, A_j\}$ of agents from the $k-1$ agents in $\mathcal{A}^\delta$.

The second key idea is to exploit one well-known property of the $s$-dimensional hyper-cube network: for every $s \geq 2$, $\mathcal{H}_s$ contains a *Hamiltonian cycle*, i.e. a simple directed cycle formed using only the edges of $\mathcal{H}_s$ and containing all $2^s$ vertices.[7] Now, suppose

$$\mathcal{S}^{(v)} = \underline{v}^{(0)}, \underline{v}^{(1)}, \ldots, \underline{v}^{(i)}, \ldots, \underline{v}^{(2^s-1)}, \underline{v}^{(0)}$$

is a Hamiltonian cycle in the hypercube $\mathcal{H}_s$ and

$$\mathcal{S}^{(w)} = \underline{w}^{(0)}, \underline{w}^{(1)}, \ldots, \underline{w}^{(i)}, \ldots, \underline{w}^{(2^s-1)}, \underline{w}^{(0)}$$

the Hamiltonian cycle in which $\underline{w}^{(i)}$ is obtained by complementing each bit in $\underline{v}^{(i)}$. As we have described in the overview of Section 2.1 we can interpret the $s$-bit label $\underline{v} = v_1 v_2 \ldots v_s$ as describing a particular subset of $\mathcal{R}^{\{i,j\}}$, i.e. that subset in which $r_k^{\{i,j\}}$ occurs if and only if $v_k = 1$. Similarly from any subset of $\mathcal{R}^{\{i,j\}}$ we may define a unique $s$-bit word. Now suppose that $P_i^{(d)}$ is the allocation held by $A_i$ in the allocation $P^{(d)}$ of $\Delta$. The deal $\delta = \langle P^{(d)}, P^{(d+1)} \rangle$ will affect $P_i^{(d)} \cap \mathcal{R}^{\{i,j\}}$ in the following way: if $i \notin \mathcal{A}^\delta$ or $j \notin \mathcal{A}^\delta$ then $P_i^{(d+1)} \cap \mathcal{R}^{\{i,j\}} = P_i^{(d)} \cap \mathcal{R}^{\{i,j\}}$ and $P_j^{(d+1)} \cap \mathcal{R}^{\{i,j\}} = P_j^{(d)} \cap \mathcal{R}^{\{i,j\}}$. Otherwise we have $\{i,j\} \subseteq \mathcal{A}^\delta$ and the (complementary) holdings $P_i^{(d)} \cap \mathcal{R}^{\{i,j\}}$ and $P_j^{(d)} \cap \mathcal{R}^{\{i,j\}}$ define (complementary) $s$-bit labels of vertices in $\mathcal{H}_s$: if these correspond to places $\langle \underline{v}^{(h)}, \underline{w}^{(h)} \rangle$ in the Hamiltonian cycles, then in $P_i^{(d+1)}$ and $P_j^{(d+1)}$ the $s$-bit labels defined from $P_i^{(d+1)} \cap \mathcal{R}^{\{i,j\}}$ and $P_j^{(d+1)} \cap \mathcal{R}^{\{i,j\}}$ produce the $s$-bit labels $\underline{v}^{(h+1)}$ and $\underline{w}^{(h+1)}$, i.e. the vertices that succeed $\underline{v}^{(h)}$ and $\underline{w}^{(h)}$ in the Hamiltonian cycles. In total, for each $j$, $A_i$ initially holds either the subset of $\mathcal{R}^{\{i,j\}}$ that maps to $\underline{v}^{(0)}$ or that maps to $\underline{w}^{(0)}$ and, at the conclusion of the $M(k-1)$-path, holds the subset that maps to $\underline{v}^{(2^s-1)}$ (or $\underline{w}^{(2^s-1)}$). The final detail is that the progression through the Hamiltonian cycles is conducted over a series of *rounds* each round comprising $k$ $M(k-1)$-deals.

We have noted that each $M(k-1)$-contract, $\langle P^{(d)}, P^{(d+1)} \rangle$ that occurs in this path $\Delta$ can be interpreted as a set of $\binom{k-1}{2}$ distinct $O$-contracts. An important property of the utility functions employed is that unless $p \geq k-1$ there will be *no individually rational* $M(p)$-contract path that realises the deal $\langle P^{(d)}, P^{(d+1)} \rangle$, i.e. the $\binom{k-1}{2}$ $O$-contract deals must occur *simultaneously* in order for the progression from $P^{(d)}$ to $P^{(d+1)}$ to be IR. Although the required deal could be realised by a sequence of $O$-contracts (or, more generally, any suitable $M(k-2)$-contract path), such realisations will *not* describe an IR contract path.

---

7. This can be shown by an easy inductive argument. For $s = 2$, the sequence $\langle 00, 01, 11, 10, 00 \rangle$ defines a Hamiltonian cycle in $\mathcal{H}_2$. Inductively assume that $\langle \alpha_1, \alpha_2, \ldots, \alpha_p, \alpha_1 \rangle$ (with $p = 2^s$) is such a cycle in $\mathcal{H}_s$ then $\langle 0\alpha_1, 1\alpha_1, 1\alpha_2, 1\alpha_{p-1}, \ldots, 1\alpha_2, 0\alpha_2 \ldots, 0\alpha_p, 0\alpha_1 \rangle$ defines a Hamiltonian cycle in $\mathcal{H}_{s+1}$.

The construction of utility functions to guarantee such behaviour provides the principal component in showing that the IR deal $\langle P^{(1)}, P^{(t)} \rangle$ cannot be realised with an IR $M(k-2)$-contract path: if $Q$ is *any* allocation for which $\langle P^{(1)}, Q \rangle$ is an $M(k-2)$-contract then $\langle P^{(1)}, Q \rangle$ is not IR.

We now proceed with the proof of Theorem 6.

*Proof.* (of Theorem 6) Fix $\mathcal{A} = \{A_0, A_1, \ldots, A_{k-1}\}$. $\mathcal{R}$ consists of $\binom{k}{2}$ pairwise disjoint sets of $s$ resources

$$\mathcal{R}^{\{i,j\}} = \{r_1^{\{i,j\}}, r_2^{\{i,j\}}, \ldots, r_s^{\{i,j\}}\}$$

For $k = 4$ and $s = 3$ these yield $\mathcal{A} = \{A_0, A_1, A_2, A_3\}$ and

$$
\begin{aligned}
\mathcal{R}^{\{0,1\}} &= \{r_1^{\{0,1\}}, r_2^{\{0,1\}}, r_3^{\{0,1\}}\} \\
\mathcal{R}^{\{0,2\}} &= \{r_1^{\{0,2\}}, r_2^{\{0,2\}}, r_3^{\{0,2\}}\} \\
\mathcal{R}^{\{0,3\}} &= \{r_1^{\{0,3\}}, r_2^{\{0,3\}}, r_3^{\{0,3\}}\} \\
\mathcal{R}^{\{1,2\}} &= \{r_1^{\{1,2\}}, r_2^{\{1,2\}}, r_3^{\{1,2\}}\} \\
\mathcal{R}^{\{1,3\}} &= \{r_1^{\{1,3\}}, r_2^{\{1,3\}}, r_3^{\{1,3\}}\} \\
\mathcal{R}^{\{2,3\}} &= \{r_1^{\{2,3\}}, r_2^{\{2,3\}}, r_3^{\{2,3\}}\}
\end{aligned}
$$

We use two ordering structures in defining the $M(k-1)$-contract path.

    a.

$$\mathcal{S}^{(v)} = \underline{v}^{(0)}, \underline{v}^{(1)}, \ldots, \underline{v}^{(i)}, \ldots, \underline{v}^{(2^s-1)}, \underline{v}^{(0)}$$

a Hamiltonian cycle in $\mathcal{H}_s$, where without loss of generality, $\underline{v}^{(0)} = 111 \ldots 11$.

    b.

$$\mathcal{S}^{(w)} = \underline{w}^{(0)}, \underline{w}^{(1)}, \ldots, \underline{w}^{(i)}, \ldots, \underline{w}^{(2^s-1)}, \underline{w}^{(0)}$$

the complementary Hamiltonian cycle to this, so that $\underline{w}^{(0)} = 000 \ldots 00$.

Thus for $k = 4$ and $s = 3$ we obtain

    a.    $\mathcal{S}^{(v)}$  =  $\langle 111, 110, 010, 011, 001, 000, 100, 101 \rangle$
    b.    $\mathcal{S}^{(w)}$  =  $\langle 000, 001, 101, 100, 110, 111, 011, 010 \rangle$

We can now describe the $M(k-1)$-contract path.

$$\Delta = \langle P^{(1)}, P^{(2)}, \ldots, P^{(t)} \rangle$$

**Initial Allocation**: $P^{(1)}$.

Define the $k \times k$ Boolean matrix, $B = [b_{i,j}]$ (with $0 \le i, j \le k-1$) by

$$
b_{i,j} = \begin{cases} \bot & \text{if} \quad i = j \\ \neg b_{j,i} & \text{if} \quad i > j \\ \neg b_{i,j-1} & \text{if} \quad i < j \end{cases}
$$

We then have for each $1 \leq i \leq k$,

$$P_i^{(1)} = \bigcup_{j=0}^{i-1} \{ R^{\{j,i\}} : b_{i,j} = \top \} \cup \bigcup_{j=i+1}^{k-1} \{ R^{\{i,j\}} : b_{i,j} = \top \}$$

Thus, in our example,

$$B = \begin{bmatrix} \bot & \top & \bot & \top \\ \bot & \bot & \top & \bot \\ \top & \bot & \bot & \top \\ \bot & \top & \bot & \bot \end{bmatrix}$$

Yielding the starting allocation

$$
\begin{array}{llll}
P_0^{(1)} = \mathcal{R}^{\{0,1\}} \cup \mathcal{R}^{\{0,3\}} & = \langle 111, 000, 111 \rangle & \subseteq & \mathcal{R}^{\{0,1\}} \cup \mathcal{R}^{\{0,2\}} \cup \mathcal{R}^{\{0,3\}} \\
P_1^{(1)} = \mathcal{R}^{\{1,2\}} & = \langle 000, 111, 000 \rangle & \subseteq & \mathcal{R}^{\{0,1\}} \cup \mathcal{R}^{\{1,2\}} \cup \mathcal{R}^{\{1,3\}} \\
P_2^{(1)} = \mathcal{R}^{\{0,2\}} \cup \mathcal{R}^{\{2,3\}} & = \langle 111, 000, 111 \rangle & \subseteq & \mathcal{R}^{\{0,2\}} \cup \mathcal{R}^{\{1,2\}} \cup \mathcal{R}^{\{2,3\}} \\
P_3^{(1)} = \mathcal{R}^{\{1,3\}} & = \langle 000, 111, 000 \rangle & \subseteq & \mathcal{R}^{\{0,3\}} \cup \mathcal{R}^{\{1,3\}} \cup \mathcal{R}^{\{2,3\}}
\end{array}
$$

The third column in $P_i^{(1)}$ indicating the 3-bit labels characterising each of the subsets of $\mathcal{R}^{\{i,j\}}$ for the three values that $j$ can assume.

**Rounds:** The initial allocation is changed over a series of *rounds*

$$Q^1, Q^2, \ldots, Q^z$$

each of which involves exactly $k$ distinct $M(k-1)$-contracts. We use $Q^{x,p}$ to indicate the allocation resulting after stage $p$ in round $x$ where $0 \leq p \leq k-1$. We note the following:

a. The initial allocation, $P^{(1)}$ will be denoted by $Q^{0,k-1}$.

b. $Q^{x,0}$ is obtained using a single $M(k-1)$-contract from $Q^{x-1,k-1}$ (when $x \geq 1$).

c. $Q^{x,p}$ is obtained using a single $M(k-1)$-contract from $Q^{x,p-1}$ (when $0 < p \leq k-1$).

Our final item of notation is that of the *cube position of $i$ with respect to $j$ in an allocation $P$*, denoted $\chi(i,j,P)$. Letting $\underline{u}$ be the $s$-bit string describing $P_i \cap \mathcal{R}^{\{i,j\}}$ in some allocation $P$, $\chi(i,j,P)$ is the *index* of $\underline{u}$ in the Hamiltonian cycle $S^{(v)}$ (when $\mathcal{R}^{\{i,j\}} \subseteq P_i^{(1)}$) or the Hamiltonian cycle $S^{(w)}$ (when $\mathcal{R}^{\{i,j\}} \subseteq P_j^{(1)}$). When $P = Q^{x,p}$ for some allocation in the sequence under construction we employ the notation $\chi(i,j,x,p)$, noting that one invariant of our path will be $\chi(i,j,x,p) = \chi(j,i,x,p)$, a property that certainly holds true of $P^{(1)} = Q^{0,k-1}$ since $\chi(i,j,0,k-1) = \chi(j,i,0,k-1) = 0$.

The sequence of allocations in $\Delta$ is built as follows. Since $Q^{1,0}$ is the immediate successor of the initial allocation $Q^{0,k-1}$, it suffices to describe how $Q^{x,p}$ is formed from $Q^{x,p-1}$ (when $p > 0$) and $Q^{x+1,0}$ from $Q^{x,k-1}$. Let $Q^{y,q}$ be the allocation to be formed from $Q^{x,p}$. The deal $\delta = \langle Q^{x,p}, Q^{y,q} \rangle$ will be an $M(k-1)$ contract in which $\mathcal{A}^\delta = \mathcal{A} \setminus \{A_q\}$. For each pair $\{i,j\} \subseteq \mathcal{A}^\delta$ we have $\chi(i,j,x,p) = \chi(j,i,x,p)$ in the allocation $Q^{x,p}$. In moving to $Q^{y,q}$ exactly one element of $\mathcal{R}^{\{i,j\}}$ is reallocated between $A_i$ and $A_j$ in such a way that in $Q^{y,q}$, $\chi(i, j, y, q) = \chi(i, j, x, p) + 1$, since $A_i$ and $A_j$ are tracing complementary Hamiltonian cycles with respect to $\mathcal{R}^{\{i,j\}}$ this ensures that $\chi(j, i, y, q) = \chi(j, i, x, p) + 1$, thereby maintaining the invariant property.

Noting that for each distinct pair $\langle i, j \rangle$, we either have $\mathcal{R}^{\{i,j\}}$ allocated to $A_i$ in $P^{(1)}$ or $\mathcal{R}^{\{i,j\}}$ allocated to $A_j$ in $P^{(1)}$, the description just outlined indicates that the allocation $P^{(d)} = Q^{x,p}$ is completely specified as follows.

The cube position, $\chi(i, j, x, p)$, satisfies,

$$\chi(i, j, x, p) = \begin{cases} 0 & \text{if} \quad x = 0 \text{ and } p = k - 1 \\ 1 + \chi(i, j, x - 1, k - 1) & \text{if} \quad x \geq 1, \ p = 0, \text{ and } p \notin \{i, j\} \\ \chi(i, j, x - 1, k - 1) & \text{if} \quad x \geq 1, \ p = 0, \text{ and } p \in \{i, j\} \\ 1 + \chi(i, j, x, p - 1) & \text{if} \quad 1 \leq p \leq k - 1, \text{ and } p \notin \{i, j\} \\ \chi(i, j, x, p - 1) & \text{if} \quad 1 \leq p \leq k - 1, \text{ and } p \in \{i, j\} \end{cases}$$

For each $i$, the subset of $\mathcal{R}^{\{i,j\}}$ that is held by $A_i$ in the allocation $Q^{x,p}$ is,

$$\underline{v}^{(\chi(i,j,x,p))} \quad \text{if} \quad \mathcal{R}^{\{i,j\}} \subseteq P_i^{(1)}$$
$$\underline{w}^{(\chi(i,j,x,p))} \quad \text{if} \quad \mathcal{R}^{\{i,j\}} \subseteq P_j^{(1)}$$

(where we recall that $s$-bit labels in the hypercube $\mathcal{H}_s$ are identified with subsets of $\mathcal{R}^{\{i,j\}}$.)

The tables below illustrates this process for our example.

| | | | $A_0$ | | | $A_1$ | | | $A_2$ | | | $A_3$ | | | $\mathcal{A}^{\langle P^{(d-1)}, P^{(d)} \rangle}$ |
|---|---|---|---|---|---|---|---|---|---|---|---|---|---|---|---|
| | | | $i\ j$ | $i\ j$ | $i\ j$ | $i\ j$ | $i\ j$ | $i\ j$ | $i\ j$ | $i\ j$ | $i\ j$ | $i\ j$ | $i\ j$ | $i\ j$ | |
| $d$ | $x$ | $p$ | 0 1 | 0 2 | 0 3 | 1 0 | 1 2 | 1 3 | 2 0 | 2 1 | 2 3 | 3 0 | 3 1 | 3 2 | |
| 1 | 0 | 3 | 111 | 000 | 111 | 000 | 111 | 000 | 111 | 000 | 111 | 000 | 111 | 000 | — |
| 2 | 1 | 0 | 111 | 000 | 111 | 000 | 110 | 001 | 111 | 001 | 110 | 000 | 110 | 001 | $\{A_1, A_2, A_3\}$ |
| 3 | 1 | 1 | 111 | 001 | 110 | 000 | 110 | 001 | 110 | 001 | 010 | 001 | 110 | 101 | $\{A_0, A_2, A_3\}$ |
| 4 | 1 | 2 | 110 | 001 | 010 | 001 | 110 | 101 | 110 | 001 | 010 | 101 | 010 | 101 | $\{A_0, A_1, A_3\}$ |
| 5 | 1 | 3 | 010 | 101 | 010 | 101 | 010 | 101 | 010 | 101 | 010 | 101 | 010 | 101 | $\{A_0, A_1, A_2\}$ |
| 6 | 2 | 0 | 010 | 101 | 011 | 101 | 011 | 100 | 010 | 100 | 011 | 101 | 011 | 100 | $\{A_1, A_2, A_3\}$ |
| 7 | 2 | 1 | 010 | 100 | 001 | 101 | 011 | 100 | 011 | 100 | 001 | 100 | 011 | 110 | $\{A_0, A_2, A_3\}$ |
| 8 | 2 | 2 | 011 | 100 | 001 | 100 | 011 | 110 | 011 | 100 | 001 | 110 | 001 | 110 | $\{A_0, A_1, A_3\}$ |
| 9 | 2 | 3 | 001 | 110 | 001 | 110 | 001 | 110 | 001 | 110 | 001 | 110 | 001 | 110 | $\{A_0, A_1, A_2\}$ |
| ⋮ | ⋮ | ⋮ | | ⋮ | | | ⋮ | | | ⋮ | | | ⋮ | | ⋮ |

**Subsets of $\mathcal{R}^{\{i,j\}}$ held by $A_i$ in $Q^{x,p}$ ($k = 4$, $s = 3$)**

|   |   |   | $A_0$ | | | $A_1$ | | | $A_2$ | | | $A_3$ | | | $\mathcal{A}^{\langle P^{(d-1)},P^{(d)}\rangle}$ |
|---|---|---|---|---|---|---|---|---|---|---|---|---|---|---|---|
| | | | $i\ j$ | $i\ j$ | $i\ j$ | $i\ j$ | $i\ j$ | $i\ j$ | $i\ j$ | $i\ j$ | $i\ j$ | $i\ j$ | $i\ j$ | $i\ j$ | |
| $d$ | $x$ | $p$ | $0\ 1$ | $0\ 2$ | $0\ 3$ | $1\ 0$ | $1\ 2$ | $1\ 3$ | $2\ 0$ | $2\ 1$ | $2\ 3$ | $3\ 0$ | $3\ 1$ | $3\ 2$ | |
| 1 | 0 | 3 | 0 | 0 | 0 | 0 | 0 | 0 | 0 | 0 | 0 | 0 | 0 | 0 | $-$ |
| 2 | 1 | 0 | 0 | 0 | 0 | 0 | 1 | 1 | 0 | 1 | 1 | 0 | 1 | 1 | $\{A_1, A_2, A_3\}$ |
| 3 | 1 | 1 | 0 | 1 | 1 | 0 | 1 | 1 | 1 | 1 | 2 | 1 | 1 | 2 | $\{A_0, A_2, A_3\}$ |
| 4 | 1 | 2 | 1 | 1 | 2 | 1 | 1 | 2 | 1 | 1 | 2 | 2 | 2 | 2 | $\{A_0, A_1, A_3\}$ |
| 5 | 1 | 3 | 2 | 2 | 2 | 2 | 2 | 2 | 2 | 2 | 2 | 2 | 2 | 2 | $\{A_0, A_1, A_2\}$ |
| 6 | 2 | 0 | 2 | 2 | 2 | 2 | 3 | 3 | 2 | 3 | 3 | 2 | 3 | 3 | $\{A_1, A_2, A_3\}$ |
| 7 | 2 | 1 | 2 | 3 | 3 | 2 | 3 | 3 | 3 | 3 | 4 | 3 | 3 | 4 | $\{A_0, A_2, A_3\}$ |
| 8 | 2 | 2 | 3 | 3 | 4 | 3 | 3 | 4 | 3 | 3 | 4 | 4 | 4 | 4 | $\{A_0, A_1, A_3\}$ |
| 9 | 2 | 3 | 4 | 4 | 4 | 4 | 4 | 4 | 4 | 4 | 4 | 4 | 4 | 4 | $\{A_0, A_1, A_2\}$ |
| $\vdots$ | $\vdots$ | $\vdots$ | $\vdots$ | | | $\vdots$ | | | $\vdots$ | | | $\vdots$ | | | $\vdots$ |

**Cube Positions** $\chi(i,j,x,p)$ $(k=4,\ s=3)$

It is certainly the case that this process of applying successive rounds of $k$ deals could be continued, however, we wish to do this only so long as it is not possible to go from some allocation $P^{(d)}$ in the sequence to another $P^{(d+r)}$ for some $r \geq 2$ via an $M(k-1)$-contract.

Now if $Q^{x,p}$ and $Q^{y,q}$ are distinct allocations generated by the process above then the deal $\delta = \langle Q^{x,p}, Q^{y,q} \rangle$ is an $M(k-1)$-contract if and only if for some $A_i$, $Q_i^{x,p} = Q_i^{y,q}$. It follows that if $\langle P^{(d)}, P^{(d+r)} \rangle$ is an $M(k-1)$-contract for some $r > 1$, then for some $i$ and all $j \neq i$, $P_i^{(d+r)} \cap \mathcal{R}^{\{i j\}} = P_i^{(d)} \cap \mathcal{R}^{\{i j\}}$.

To determine the minimum value of $r > 1$ with which $P_i^{(d+r)} = P_i^{(d)}$, we observe that without loss of generality we need consider only the case $d = i = 0$, i.e. we determine the minimum number of deals before $P_0^{(1)}$ reappears. First note that in each round, $Q^x$, if $\chi(0,j,x-1,k-1) = p$ then $\chi(0,j,x,k-1) = p+k-2$, i.e. each round advances the cube position $k-2$ places: $\chi(0,j,x-1,k-1) = \chi(0,j,x,0)$ and $\chi(0,j,x,j) = \chi(0,j,x,j-1)$. We can also observe that $P_0^{(1)} = Q_0^{0,k-1} \neq Q_0^{x,p}$ for any $p$ with $0 < p < k-1$, since

$$\chi(0,1,x,p) = \chi(0,2,x,p) = \ldots = \chi(0,k-1,x,p)$$

only in the cases $p = 0$ and $p = k-1$. It follows that our value $r > 1$ must be of the form $qk$ where $q$ must be such that $q(k-2)$ is an *exact multiple* of $2^s$. From this observation we see that,

$$\min\{\ r > 1 : P_0^{(1)} = P_0^{(1+r)}\ \} \ = \ \min\{\ qk : q(k-2) \text{ is a multiple of } 2^s\}$$

Now, if $k$ is *odd* then $q = 2^s$ is the minimal such value, so that $r = k2^s$. If $k$ is even then it may be uniquely written in the form $z2^l + 2$ where $z$ is *odd* so giving $q$ as 1 (if $l \geq s$) or $2^{s-l}$ (if $l \leq s$), so that these give $r = k$ and $r = z2^s + 2^{s-l+1}$, e.g. for $k = 4$ and $s = 3$, we get $k = 1 \times 2^1 + 2$ so that $r = 2^3 + 2^{3-1+1} = 16$ and in our example $P_0^{(1)} = P_0^{(17)}$ may be easily verified. In total,

$$r \geq \begin{cases} k2^s & \text{if}\quad k \text{ is odd} \\ k & \text{if}\quad k = z2^l + 2,\ z \text{ is odd, and } l \geq s \\ 2^s & \text{if}\quad k = z2^l + 2,\ z \text{ is odd and } l \leq s \end{cases}$$

All of which immediately give $r \geq 2^s$ (in the second case $k \geq 2^s$, so the inequality holds trivially), and thus we can continue the chain of $M(k-1)$ contracts for at least $2^s$ moves. Recalling that $m = s\binom{k}{2}$, this gives the length of the $M(k-1)$-contract path

$$\Delta = \langle P^{(1)}, P^{(2)}, \ldots, P^{(t)} \rangle$$

written in terms of $m$ and $k$ as at least[8]

$$2^{m/\binom{k}{2}} - 1 = 2^{\frac{2m}{k(k-1)}} - 1$$

It remains to define appropriate utility functions $\mathcal{U} = \langle u_0, \ldots, u_{k-1} \rangle$ in order to ensure that $\Delta$ is the unique IR $M(k-1)$-contract path realising the IR $M(k)$-deal $\langle P^{(1)}, P^{(t)} \rangle$. In defining $\mathcal{U}$ it will be convenient to denote $\Delta$ as the path

$$\Delta = \langle Q^{0,k-1}, Q^{1,0}, Q^{1,1}, \ldots, Q^{1,k-1}, \ldots, Q^{x,p}, \ldots, Q^{r,k-1} \rangle$$

and, since $rk \geq 2^s$, we may without loss of generality, focus on the first $2^s$ allocations in this contract path.

Recalling that $\chi(i, j, x, p)$ is the index of the $s$-bit label $\underline{u}$ corresponding to $Q_i^{x,p} \cap \mathcal{R}^{\{i,j\}}$ in the relevant Hamiltonian cycle – i.e. $\mathcal{S}^{(v)}$ if $\mathcal{R}^{\{i,j\}} \subseteq Q_i^{0,k}$, $\mathcal{S}^{(w)}$ if $\mathcal{R}^{\{i,j\}} \subseteq Q_j^{0,k-1}$ – we note the following properties of the sequence of allocations defined by $\Delta$ that hold for each distinct $i$ and $j$.

P1. $\forall \, x, p \; \chi(i, j, x, p) = \chi(j, i, x, p)$

P2. If $Q^{y,q}$ is the immediate successor of $Q^{x,p}$ in $\Delta$ then $\chi(i, j, y, q) \leq \chi(i, j, x, p) + 1$ with equality if and only if $q \notin \{i, j\}$.

P3. $\forall \, i', j'$ with $0 \leq i', j' \leq k-1$, $\chi(i, j, x, k-1) = \chi(i', j', x, k-1)$.

The first two properties have already been established in our description of $\Delta$. The third follows from the observation that within each round $Q^x$, each cube position is advanced by exactly $k-2$ in progressing from $Q^{x-1,0}$ to $Q^{x,k-1}$.

The utility function $u_i$ is now given, for $S \subseteq \mathcal{R}_m$, by

$$u_i(S) = \begin{cases} \sum_{j \neq i} \chi(i, j, x, p) & \text{if } S = Q_i^{x,p} \text{ for some } 0 \leq x \leq r, \, 0 \leq p \leq k-1 \\ -2^{km} & \text{otherwise} \end{cases}$$

We claim that, with these choices,

$$\Delta = \langle Q^{0,k-1}, Q^{1,0}, Q^{1,1}, \ldots, Q^{1,k-1}, \ldots, Q^{x,p}, \ldots, Q^{r,k-1} \rangle$$

is the unique IR $M(k-1)$-contract path realising the IR $M(k)$-deal $\langle Q^{0,k-1}, Q^{r,k-1} \rangle$. Certainly, $\Delta$ *is* an IR $M(k-1)$-contract path: each deal $\delta = \langle Q^{x,p}, Q^{y,q} \rangle$ on this path has $|\mathcal{A}^\delta| = k-1$ and since for each agent $A_i$ in $\mathcal{A}^\delta = \mathcal{A} \setminus \{A_q\}$ the utility of $Q_i^{y,q}$ has increased

---

8. We omit the rounding operation $\lfloor \ldots \rfloor$ in the exponent, which is significant only if $m$ is not an exact multiple of $\binom{k}{2}$, in which event the device described in our overview of the proof is applied.

by exactly $k-2$, i.e. each cube position of $i$ with respect to $j$ whenever $q \notin \{i,j\}$ has increased, it follows that $\sigma_u(Q^{y,q}) > \sigma_u(Q^{x,p})$ and hence $\langle Q^{x,p}, Q^{y,q} \rangle$ is IR.

We now show that $\Delta$ is the *unique* IR $M(k-1)$-contract path continuation of $Q^{0,k-1}$ Suppose $\delta = \langle Q^{x,p}, P \rangle$ is a deal that deviates from the contract path $\Delta$ (having followed it through to the allocation $Q^{x,p}$). Certainly both of the following must hold of $P$: for each $i$, $P_i \subseteq \cup_{j \neq i} \mathcal{R}^{\{i,j\}}$; and there is a $k$-tuple of pairs $\langle (x_0, p_0), \ldots, (x_{k-1}, p_{k-1}) \rangle$ with which $P_i = Q_i^{x_i, p_i}$, for if either fail to be the case for some $i$, then $u_i(P_i) = -2^{km}$ with the consequent effect that $\sigma_u(P) < 0$ and thence not IR. Now, if $Q^{y,q}$ is the allocation that would succeed $Q^{x,p}$ in $\Delta$ then $P \neq Q^{y,q}$, and thus for at least one agent, $Q_i^{x_i, p_i} \neq Q_i^{y,q}$. It cannot be the case that $Q_i^{x_i, p_i}$ corresponds to an allocation occurring *strictly later* than $Q_i^{y,q}$ in $\Delta$ since such allocations could not be realised by an $M(k-1)$-contract. In addition, since $P_i = Q_i^{x_i, p_i}$ it must be the case that $|\mathcal{A}^\delta| = k-1$ since exactly $k-1$ cube positions in the holding of $A_i$ must change. It follows that there are only two possibilities for $(y_i, p_i)$: $P_i$ reverts to the allocation immediately preceding $Q_i^{x,p}$ or advances to the holding $Q_i^{y,q}$. It now suffices to observe that a deal in which some agents satisfy the first of these while the remainder proceed in accordance with the second either does not give rise to a valid allocation or cannot be realised by an $M(k-1)$-contract. On the other hand if $P$ corresponds to the allocation preceding $Q^{x,p}$ then $\delta$ is not IR. We deduce, therefore, that the *only* IR $M(k-1)$ deal that is consistent with $Q^{x,p}$ is that prescribed by $Q^{y,q}$.

This completes the analysis needed for the proof of part (b) of the theorem. It is clear that since the system contains only $k$ agents, any deal $\langle P, Q \rangle$ can be effected with a single $M(k)$-contract, thereby establishing part (a). For part (c) – that the IR deal $\langle P^{(1)}, P^{(t)} \rangle$ cannot be realised using an *individually rational* $M(k-2)$-contract path, it suffices to observe that since the class of IR $M(k-2)$-contracts are a subset of the class of IR $M(k-1)$-contracts, were it the case that an IR $M(k-2)$-contract path existed to implement $\langle P^{(1)}, P^{(t)} \rangle$, this would imply that $\Delta$ was not the *unique* IR $M(k-1)$-contract path. We have, however, proved that $\Delta$ *is* unique, and part (c) of the theorem follows. □

We obtain a similar development of Corollary 1 in

**Corollary 3** *For all $k \geq 3$, $n \geq k$, $m \geq \binom{k}{2}$ and each of the cases below,*

a. $\Phi_k(\delta)$ *holds if and only if $\delta$ is a cooperatively rational $M(k)$-contract.*
   $\Psi(\delta)$ *holds if and only if $\delta$ is cooperatively rational.*

b. $\Phi_k(\delta)$ *holds if and only if $\delta$ is $\delta$ is an equitable $M(k)$-contract.*
   $\Psi(\delta)$ *holds if and only if $\delta$ is is equitable.*

*there is a resource allocation setting $\langle \mathcal{A}, \mathcal{R}, \mathcal{U} \rangle$ and a $\Psi$-deal $\delta = \langle P, Q \rangle$ for which*

$$
\begin{array}{llcr}
L^{\text{opt}}(\delta, \langle \mathcal{A}, \mathcal{R}, \mathcal{U} \rangle, \Phi_k) & = & 1 & (a) \\
L^{\text{opt}}(\delta, \langle \mathcal{A}, \mathcal{R}, \mathcal{U} \rangle, \Phi_{k-1}) & \geq & 2^{\lfloor 2m/k(k-1) \rfloor} - 1 & (b) \\
L^{\text{opt}}(\delta, \langle \mathcal{A}, \mathcal{R}, \mathcal{U} \rangle, \Phi_{k-2}) & & \text{is undefined} & (c)
\end{array}
$$

*Proof.* As with the proof of Corollary 1 in relation to Theorem 3, in each case we employ the contract path from the proof of Theorem 6, varying the definition of $\mathcal{U} = \langle u_1, u_2, \ldots, u_k \rangle$ in order to establish each result. Thus let

$$
\begin{array}{lll}
\Delta_m & = & \langle P^{(1)}, P^{(2)}, \ldots, P^{(r)}, \ldots, P^{(t)} \rangle \\
& = & \langle Q^{0,k-1}, Q^{1,0}, \ldots, Q^{x,p}, \ldots, Q^{z,r} \rangle
\end{array}
$$

be the $M(k-1)$-contract path realising the $M(k)$-deal $\langle P^{(1)}, P^{(t)} \rangle$ described in the proof of Theorem 6, this path having length $t \geq 2^{\lfloor 2m/k(k-1) \rfloor} - 1$.

a. The utility functions $\mathcal{U} = \langle u_0, \ldots, u_{k-1} \rangle$ of Theorem 6 ensure that $\langle P^{(1)}, P^{(t)} \rangle$ is cooperatively rational and that $\Delta_m$ is a cooperatively rational $M(k-1)$-contract path realising $\langle P^{(1)}, P^{(t)} \rangle$: the utility held by $A_i$ never decreases in value and there is at least one agent (in fact exactly $k-1$) whose utility increases in value. Furthermore $\Delta_m$ is the unique cooperatively rational $M(k-1)$-contract path realising $\langle P^{(1)}, P^{(t)} \rangle$ since, by the same argument used in Theorem 6, any deviation will result in some agent suffering a loss of utility.

b. Set the utility functions $\mathcal{U} = \langle u_0, \ldots, u_{k-1} \rangle$ as,

$$
u_i(S) = \begin{cases}
-1 & \text{if} \quad S \neq Q_i^{x,p} \text{ for any } Q^{x,p} \in \Delta_m \\
xk^2 + k - i & \text{if} \quad S = Q_i^{x,k-1} \\
(x-1)k^2 + k + p & \text{if} \quad S = Q_0^{x,p}, \ p < k-1 \text{ and } i = 0 \\
(x-1)k^2 + k - i + p + 1 & \text{if} \quad S = Q_i^{x,p}, \ p < i-1 \text{ and } i \neq 0. \\
xk^2 + 1 & \text{if} \quad S = Q_i^{x,i-1} = Q_i^{x,i} \text{ and } i \neq 0. \\
xk^2 + 1 + p - i & \text{if} \quad S = Q_i^{x,p}, \ p > i \text{ and } i \neq 0
\end{cases}
$$

To see that these choices admit $\Delta_m$ as an equitable $M(k-1)$-contract path realising the equitable deal $\langle Q^{0,k-1}, Q^{z,r} \rangle$, we first note that

$$
\min_{0 \leq i \leq k-1} \ \{ u_i(Q_i^{z,r}) \} \ > \ 1 \ = \ \min_{0 \leq i \leq k-1} \ \{ u_i(Q_i^{0,k-1}) \}
$$

thus, $\langle Q^{0,k-1}, Q^{z,r} \rangle$ is indeed equitable. Consider any deal $\delta = \langle Q^{x,p}, Q^{y,q} \rangle$ occurring within $\Delta_m$. It suffices to show that

$$
\min_{0 \leq i \leq k-1} \ \{ u_i(Q_i^{x,p}) \} \ \neq \ u_q(Q_q^{x,p})
$$

since $A_q \notin \mathcal{A}^\delta$, and for all other agents $u_i(Q_i^{y,q}) > u_i(Q_i^{x,p})$. We have two possibilities: $q = 0$ (in which case $p = k-1$ and $y = x + 1$); $q > 0$ (in which case $p = q - 1$). Consider the first of these: $u_0(Q_0^{x,k-1}) = xk^2 + k$, however,

$$
\min\{u_i(Q_i^{x,k-1})\} = xk^2 + 1 = u_{k-1}(Q_{k-1}^{x,k-1})
$$

and hence every deal $\langle Q^{x,k-1}, Q^{x+1,0} \rangle$ forming part of $\Delta_m$ is equitable.

In the remaining case, $u_q(Q_q^{x,q-1}) = xk^2 + 1$ and

$$
\begin{aligned}
\min\{u_i(Q_i^{x,q-1})\} \quad &\leq \quad u_0(Q_0^{x,q-1}) \\
&= \quad (x-1)k^2 + k + q - 1 \\
&< \quad xk^2 - (k^2 - 2k + 1) \\
&= \quad xk^2 - (k-1)^2 \\
&< \quad xk^2 + 1 \\
&= \quad u_q(Q_q^{x,q-1})
\end{aligned}
$$

and thus the remaining deals $\langle Q^{x,q-1}, Q^{x,q} \rangle$ within $\Delta_m$ are equitable. By a similar argument to that employed in Theorem 6 it follows that $\Delta_m$ is the unique equitable $M(k-1)$-contract path realising $\langle Q^{0,k-1}, Q^{z,r} \rangle$.

$\square$

## Monotone Utility Functions and $M(k)$-contract paths

The device used to develop Theorem 3 to obtain the path of Theorem 4 can be applied to the rather more intricate construction of Theorem 6, thereby allowing exponential lower bounds on $\rho_{\mathrm{mono}}^{\max}(n, m, \Phi_k, \Psi)$ to be derived. We will merely outline the approach rather than present a detailed technical exposition. We recall that it became relatively straightforward to define suitable monotone utility functions once it was ensured that the subset sizes of interest – i.e. those for allocations arising in the $O$-contract path – were forced to fall into a quite restricted range. The main difficulty that arises in applying similar methods to the path $\Delta$ of Theorem 6 is the following: in the proof of Theorem 4 we consider two agents so that converting $\Delta_s$ from a setting with $s$ resources in Theorem 3 to $ext(\Delta_s)$ with $2s$ resources in Theorem 4 is achieved by combining "complementary" allocations, i.e. $\alpha \subseteq \mathcal{R}_s$ with $\overline{\alpha} \subseteq \mathcal{T}_s$. We can exploit two facts, however, to develop a path $multi(\Delta)$ for which monotone utility functions could be defined: the resource set $\mathcal{R}_m$ in Theorem 6 consists of $\binom{k}{2}$ disjoint sets of size $s$; and any deal $\delta$ on the path $\Delta$ involves a reallocation of $\mathcal{R}^{\{i,j\}}$ between $A_i$ and $A_j$ when $\{i, j\} \subseteq \mathcal{A}^{\delta}$. Thus letting $\mathcal{T}_m$ be formed by $\binom{k}{2}$ disjoint sets, $\mathcal{T}^{\{i,j\}}$ each of size $s$, suppose that $P_i^{(d)}$ is described by

$$\alpha_{i,0}^{(d)} \; \alpha_{i,1}^{(d)} \; \cdots \; \alpha_{i,i-1}^{(d)} \; \alpha_{i,i+1}^{(d)} \; \cdots \; \alpha_{i,k-1}^{(d)}$$

with $\alpha_{i,j}^{(d)}$ the $s$-bit label corresponding to the subset of $R^{\{i,j\}}$ that is held by $A_i$ in $P^{(d)}$. Consider the sequence of allocations,

$$multi(\Delta) \; = \; \langle C^{(1)}, C^{(2)}, \ldots, C^{(t)} \rangle$$

in a resource allocation setting have $k$ agents and $2m$ resources – $\mathcal{R}_m \cup \mathcal{T}_m$ for which $C_i^{(d)}$ is characterised by

$$\beta_{i,0}^{(d)} \; \beta_{i,1}^{(d)} \; \cdots \; \beta_{i,i-1}^{(d)} \; \beta_{i,i+1}^{(d)} \; \cdots \; \beta_{i,k-1}^{(d)}$$

In this, $\beta_{i,j}^{(d)}$, indicates the subset of $\mathcal{R}^{\{i,j\}} \cup \mathcal{T}^{\{i,j\}}$ described by the $2s$-bit label,

$$\beta_{i,j}^{(d)} \; = \; \alpha_{i,j}^{(d)} \; \overline{\alpha_{i,j}^{(d)}}$$

i.e. $\alpha_{i,j}^{(d)}$ selects a subset of $\mathcal{R}^{\{i,j\}}$ while $\overline{\alpha_{i,j}^{(d)}}$ a subset of $\mathcal{T}^{\{i,j\}}$.

It is immediate from this construction that for each allocation $C^{(d)}$ in $multi(\Delta)$ and each $A_i$, it is always the case that $|C_i^{(d)}| = (k-1)s$. It follows, therefore, that the only subsets that are relevant to the definition of *monotone* utility functions with which an analogous result to Theorem 6 for the path $multi(\Delta)$ could be derived, are those of size $(k-1)s$: if $S \subseteq \mathcal{R}_m \cup \mathcal{T}_m$ has $|S| < (k-1)s$, we can fix $u_i(S)$ as a small enough negative value; similarly if $|S| > (k-1)s$ then $u_i(S)$ can be set to a large enough positive value.[9]

Our description in the preceding paragraphs, can be summarised in the following result, whose proof is omitted: extending the outline given above to a formal lower bound proof, is largely a technical exercise employing much of the analysis already introduced, and since nothing signifcantly new is required for such an analysis we shall not give a detailed presentation of it.

---

9. It is worth noting that the "interpolation" stage used in Theorem 4 is not needed in forming $multi(\Delta)$: the deal $\langle C^{(d)}, C^{(d+1)} \rangle$ is an $M(k-1)$-contract. We recall that in going from $\Delta_s$ of Theorem 3 to $ext(\Delta_s)$ the intermediate stage – $double(\Delta_s)$ – was not an $O$-contract path.

**Theorem 7** *Let $\Phi_k(P, Q)$ be the predicate which holds whenever $\langle P, Q \rangle$ is an IR $M(k)$-contract. For all $k \geq 3$, $n \geq k$ and $m \geq 2 \binom{k}{2}$, there is a resource allocation setting $\langle \mathcal{A}, \mathcal{R}, \mathcal{U} \rangle$ in which every $u \in \mathcal{U}$ is monotone, and an IR deal $\delta = \langle P, Q \rangle$ for which,*

$$
\begin{array}{lccl}
L^{\text{opt}}(\delta, \langle \mathcal{A}, \mathcal{R}, \mathcal{U} \rangle, \Phi_k) & = & 1 & (a) \\
L^{\text{opt}}(\delta, \langle \mathcal{A}, \mathcal{R}, \mathcal{U} \rangle, \Phi_{k-1}) & \geq & 2^{\lfloor m/k(k-1) \rfloor} - 1 & (b) \\
L^{\text{opt}}(\delta, \langle \mathcal{A}, \mathcal{R}, \mathcal{U} \rangle, \Phi_{k-2}) & & \text{is undefined} & (c)
\end{array}
$$

## 4. Related Work

The principal focus of this article has considered a property of contract paths realising rational reallocations $\langle P, Q \rangle$ when the constituent deals are required to conform to a structural restriction and satisfy a rationality constraint. In Section 2 the structural restriction limited deals to those involving a single resource, i.e. $O$-contracts. For the rationality constraint forcing deals strictly to improve utilitarian social welfare, i.e. to be individually rational (IR) we have the following properties.

a. There are resource allocation settings $\langle \mathcal{A}, \mathcal{R}, \mathcal{U} \rangle$ within which there are IR reallocations $\langle P, Q \rangle$ that *cannot* be realised by a sequence of IR $O$-contracts. (Sandholm, 1998, Proposition 2)

b. Every IR reallocation, $\langle P, Q \rangle$, that *can* be realised by an IR $O$-contract path, can be realised by an IR $O$-contract path of length at most $n^m - (n-1)m$. (Sandholm, 1998, Proposition 2)

c. Given $\langle \mathcal{A}, \mathcal{R}, \mathcal{U} \rangle$ together with an IR reallocation $\langle P, Q \rangle$ the problem of deciding if $\langle P, Q \rangle$ can be implemented by an IR $O$-contract path is NP–hard, even if $|\mathcal{A}| = 2$ and both utility functions are monotone. (Dunne et al., 2003, Theorem 11).

d. There are resource allocation settings $\langle \mathcal{A}, \mathcal{R}, \mathcal{U} \rangle$ within which there are IR reallocations $\langle P, Q \rangle$ that *can* be realised by an IR $O$-contract path, but with any such path having length exponential in $m$. This holds even in the case $|\mathcal{A}| = 2$ and both utility functions are monotone. (Theorem 3 and Theorem 4 of Section 2)

In a recent article Endriss and Maudet (2004a) analyse contract path length also considering $O$-contracts with various rationality constraints. Although the approach is from a rather different perspective, the central question addressed – "How many rational deals are required to reach an optimal allocation?", (Endriss & Maudet, 2004a, Table 1, p. 629) – is closely related to the issues discussed above. One significant difference in the analysis of rational $O$-contracts from Sandholm's (1998) treatment and the results in Section 2 is that in (Endriss & Maudet, 2004a) the utility functions are restricted so that *every* rational reallocation $\langle P, Q \rangle$ *can* be realised by a rational $O$-contract path. The two main restrictions examined are requiring utility functions to be *additive*, i.e. for every $S \subseteq \mathcal{R}$, $u(S) = \sum_{r \in S} u(r)$;

and, requiring the value returned to be either 0 or 1, so-called $0 - 1$ utility functions. Additive utility functions are considered in the case of IR $O$-contracts (Endriss & Maudet, 2004a, Theorems 3, 9), whereas $0-1$ utility functions for cooperatively rational $O$-contracts (Endriss & Maudet, 2004a, Theorems 4, 11). Using $\rho_{\mathrm{add}}^{\max}(n, m, \Phi, \Psi)$ and $\rho_{0-1}^{\max}(n, m, \Phi, \Psi)$ to denote the functions introduced in Definition 6 where all utility functions are additive (respectively $0-1$), cf. the definition of $\rho_{\mathrm{mono}}^{\max}$, then with $\Phi_1(P, Q)$ holding if $\langle P, Q \rangle$ is an IR $O$-contract; $\Phi_2(P, Q)$ holding if $\langle P, Q \rangle$ is a cooperatively rational $O$-contract and $\Psi(P, Q)$ true when $\langle P, Q \rangle$ is IR, we may formulate Theorems 9 and 11 of (Endriss & Maudet, 2004a) in terms of the framework used in Definition 6, as

$$\rho_{\mathrm{add}}^{\max}(n, m, \Phi_1, \Psi) = m \qquad (Endriss \ \& \ Maudet, 2004a, \ Theorem \ 9)$$
$$\rho_{0-1}^{\max}(n, m, \Phi_2, \Psi) = m \qquad (Endriss \ \& \ Maudet, 2004a, \ Theorem \ 11)$$

We can, of course, equally couch Theorems 3 and 4 of Section 2 in terms of the "shortest-path" convention adopted in (Endriss & Maudet, 2004a), provided that the domains of utility and reallocation instances are restricted to those for which an appropriate $O$-contract path exists. Thus, we can obtain the following development of (Endriss & Maudet, 2004a, Table 1) in the case of $O$-contracts.

| Utility Functions | Additive | 0-1 | Unrestricted | Monotone | Unrestricted | Monotone |
|---|---|---|---|---|---|---|
| Rationality | IR | CR | IR | IR | CR | CR |
| Shortest Path | $m$ | $m$ | $\Omega(2^m)$ | $\Omega(2^{m/2})$ | $\Omega(2^m)$ | $\Omega(2^{m/2})$ |
| Complete | Yes | Yes | No | No | No | No |

Table 2: How many $O$-contract rational deals are required to reach an allocation? Extension of Table 1 from (Endriss & Maudet, 2004a, p. 629)

## 5. Conclusions and Further Work

Our aim in this article has been to develop the earlier studies of Sandholm (1998) concerning the scope and limits of particular "practical" contract forms. While Sandholm (1998) has established that insisting on individual rationality in addition to the structural restriction prescribed by $O$-contracts leads to scenarios which are incomplete (in the sense that there are individually rational deals that cannot be realised by individually rational $O$-contracts) our focus has been with respect to deals which can be realised by restricted contract paths, with the intention of determining to what extent the combination of structural and rationality conditions increases the number of deals required. We have shown that, using a number of natural definitions of rationality, for settings involving $m$ resources, *rational* $O$-contract paths of length $\Omega(2^m)$ are needed, whereas without the rationality restriction on individual deals, at most $m$ $O$-contracts suffice to realise any deal. We have also considered a class of deals – $M(k)$-contracts – that were not examined in (Sandholm, 1998), establishing for these cases that, when particular rationality conditions are imposed, $M(k-1)$-contract paths of length $\Omega(2^{2m/k^2})$ are needed to realise a deal that can be achieved by a single $M(k)$-contract.

We note that our analyses have primarily been focused on worst-case lower bounds on path length when appropriate paths exist, and as such there are several questions of practical interest that merit further discussion. It may be noted that the path structures and associated utility functions are rather artificial, being directed to attaining a path of a specific length meeting a given rationality criterion. We have seen, however, in Theorems 4 and 5 as outlined in our discussion concluding Section 3 that the issue of exponential length contract paths continues to arise even when we require the utility functions to satisfy a monotonicity condition. We can identify two classes of open question that arise from these results.

Firstly, focusing on IR $O$-contract paths, it would be of interest to identify "natural" restrictions on utility functions which would ensure that, *if* a deal $\langle P, Q \rangle$ can be implemented by an IR $O$-contract path, then it can be realised by one whose length is polynomially bounded in $m$, e.g. such as additivity mentioned in the preceding section. We can interpret Theorem 4, as indicating that monotonicity does *not* guarantee "short" IR contract paths. We note, however, that there *are* some restrictions that suffice. To use a rather trivial example, if the number of *distinct* values that $\sigma_u$ can assume is at most $m^p$ for some *constant* $p$ then no IR $O$-contract path can have length exceeding $m^p$: successive deals must strictly increase $\sigma_u$ and if this can take at most $K$ different values then no IR contract path can have length exceeding $K$. As well as being of practical interest, classes of utility function with the property being considered would also be of some interest regarding one complexity issue. The result proved in (Dunne et al., 2003) establishing that deciding if an IR $O$-contract path exists is NP-hard, gives a *lower bound* on the computational complexity of this problem. At present, no (non-trivial) *upper bound* on this problem's complexity has been demonstrated. Our results in Theorems 3 and 4 indicate that *if* this decision problem is in NP (thus its complexity would be NP–complete rather than NP–hard) then the required polynomial length existence certificate *may* have to be something other than the path itself.[10] We note that the proof of NP–hardness in (Dunne et al., 2003) constructs an instance in which $\sigma_u$ can take at most $O(m)$ distinct values: thus, from our example of a restriction ensuring that if such are present then IR $O$-contract paths are "short", this result of (Dunne et al., 2003) indicates that the question of *deciding* their existence might remain computationally hard.

Considering restrictions on the form of utility functions is one approach that could be taken regarding finding "tractable" cases. An alternative would be to gain some insight into what the "average" path length is likely to be. In attempting to address this question, however, a number of challenging issues arise. The most immediate of these concerns, of course, the notion of modeling a distribution on utility function given our definitions of rationality in terms of the value agents attach to their resource holdings. In principle an average-case analysis of scenarios involving exactly *two* agents could be carried out in purely graph-theoretic terms, i.e. without the complication of considering utility functions directly. It is unclear, however, whether such a graph-theoretic analysis obviating the need for consideration of literal utility functions, can be extended beyond settings involving exactly two agents. One difficulty arising with three or more agents is that our utility functions have no allocative externalities, i.e. given an allocation $\langle X, Y, Z \rangle$ to three agents, $u_1(X)$ is unchanged should $Y \cup Z$ be redistributed among $A_2$ and $A_3$.[11]

---

10. The use of "may" rather than "must" is needed because of the convention for representing utility functions employed in (Dunne et al., 2003).

As one final set of issues that may merit further study we raise the following. In our constructions, the individual deals on a contract path must satisfy both a *structural* condition (be an $O$-contract or involve at most $k$ agents), and a *rationality* constraint. Focusing on $O$-contracts we have the following extremes: from (Sandholm, 1998), at most $m$ $O$-contracts suffice to realise any rational deal; from our results above, $\Omega(2^m)$ *rational* $O$-contracts are needed to realise some rational deals. There are a number of mechanisms we can employ to relax the condition that every single deal be an $O$-contract and be rational. For example, allow a path to contain some number of deals which are not $O$-contracts (but must still be IR) or insist that all deals are $O$-contracts but allow some to be irrational. Thus, in the latter case, if we go to the extent of allowing up to $m$ irrational $O$-contracts, then any rational deal can be realised efficiently. It would be of some interest to examine issues such as the effect of allowing a *constant* number, $t$, of irrational deals and questions such as whether there are situations in which $t$ irrational contracts yield a 'short' contract path but $t - 1$ force one of exponential length. Of particular interest, from an application viewpoint, is the following: define a $(\gamma(m), O)$-path as an $O$-contract path containing at most $\gamma(m)$ $O$-contracts which are not individually rational. We know that if $\gamma(m) = 0$ then individually rational $(0, O)$-paths are not complete with respect to individually rational deals; similarly if $\gamma(m) = m$ then $(m, O)$-paths *are* complete with respect to individually rational deals. A question of some interest would be to establish if there is some $\gamma(m) = o(m)$ for which $(\gamma(m), O)$-paths are complete with respect to individually rational deals *and* with the maximum length of such a contract path bounded by a polynomial function of $m$.

## Acknowledgements

The author thanks the reviewers of an earlier version of this article for their valuable comments and suggestions which have contributed significantly to its content and organisation. The work reported in this article was carried out under the support of EPSRC Grant GR/R60836/01.

---

11. A very preliminary investigation of complexity-theoretic questions arising in settings with allocative externalities is presented in (Dunne, 2004) where these are referred to as "context–dependent": such utility functions appear to have been neglected in the computational and algorithmic analysis of resource allocation problems, although the idea is well-known to game-theoretic models in economics from which the term "allocative externality" originates.